\documentclass[reprint]{revtex4-2}

\usepackage{graphicx}
\usepackage{bm}
\usepackage{float}
\usepackage{subcaption}
\usepackage{amsmath}
\usepackage{textcomp}
\usepackage{gensymb}
\usepackage{xcolor}
\usepackage{soul}
\usepackage{upgreek}

\begin{document}

\title{Frozen mode in an asymmetric serpentine optical waveguide}

\author{A. Herrero-Parareda}
 \altaffiliation{Department of Electrical Engineering and Computer Science, University of California, Irvine, Irvine, CA, 92617.}
\author{I. Vitebskiy}%
\altaffiliation{Air Force Research Laboratory, Sensors Directorate, Wright-Patterson AFB, OH 45433, USA.}
\author{J. Scheuer}
\altaffiliation{School of Electrical Engineering Tel Aviv University, Ramat Aviv, Tel-Aviv 69978, Israel.}
\author{F. Capolino}
\altaffiliation{Department of Electrical Engineering and Computer Science, University of California, Irvine, Irvine, CA, 92617.}
\email{f.capolino@uci.edu}

\date{\today}

\begin{abstract}
The existence of a frozen mode in a periodic serpentine waveguide with broken longitudinal symmetry is demonstrated numerically. The frozen mode is associated with a stationary inflection point (SIP) of the Bloch dispersion relation, where three Bloch eigenmodes collapse on each other, as it is an exceptional point of order three. The frozen mode regime is characterized by vanishing group velocity and enhanced field amplitude, which can be very attractive in various applications including dispersion engineering, lasers, and delay lines. Useful and simple design equations that lead to realization of the frozen mode by adjusting a few parameters are derived. The trend in group delay and quality factor with waveguide length that is peculiar of the frozen mode is shown. The symmetry conditions for the existence of exceptional points of degeneracy associated with the frozen mode are also discussed.
\end{abstract}


\maketitle
\section{Introduction}
\label{ch:intro}

The confinement and slowing down of light in photonic structures has gained interest in the past two decades due to its growing feasibility and possible applications.
Of particular interest is the excitation of the frozen mode regime \cite{figotin_slow_2011}, where the wave transmitted inside a waveguide or in a supporting medium exhibits both a vanishing group velocity and an enhanced amplitude \cite{figotin_slow_2006}. The frozen mode regime is associated with a stationary inflection point (SIP) of the Bloch dispersion relation $\omega(k)$, where $\partial \omega / \partial k = 0$ and $\partial^2\omega/\partial k^2 = 0$ at $k = k_s$, where $k_s$ is the SIP wavenumber. In this paper, we focus on SIPs because of their diverse potential applications: loss-induced transparency, unidirectional invisibility, lasing mode selection, lasing revivals and suppression, directional lasing, hypersensitive sensors, etc \cite{thomas_giant_2016}. The SIP scenario is also interesting and attractive because the frozen mode regime can be observed over a wide frequency range, ranging from RF  \cite{nada_frozen_2021, mumcu_RF_2005}, to optical frequencies \cite{nada_theory_2017, paul_frozen_2021, ScheuerSumetsky}.
Moreover, third order exceptional points of degeneracy (EPDs) have been found in a diverse range of structures:  loss-gain balanced coupled mode structures, such as PT-symmetric systems with glide symmetry \cite{ramezani_unidirectional_2010, yazdi_SIP_2021}, SIPs are found in periodic lossless and gainless coupled mode structures \cite{nada_theory_2017, li_frozen_2017}, periodic lossless and gainless gratings \cite{gutman_stationary_2011} and photonic crystals \cite{figotin_slow_2006}.
Furthermore, SIPs have been found in nonreciprocal structures, as shown in \cite{ramezani_unidirectional_2014, Stephanson_ferromagnetic_2008, figotin_nonreciprocal_2004}, where the system becomes unidirectional near the SIP frequency.

\begin{figure}[H]
\centering
\includegraphics[width = 0.48\textwidth]{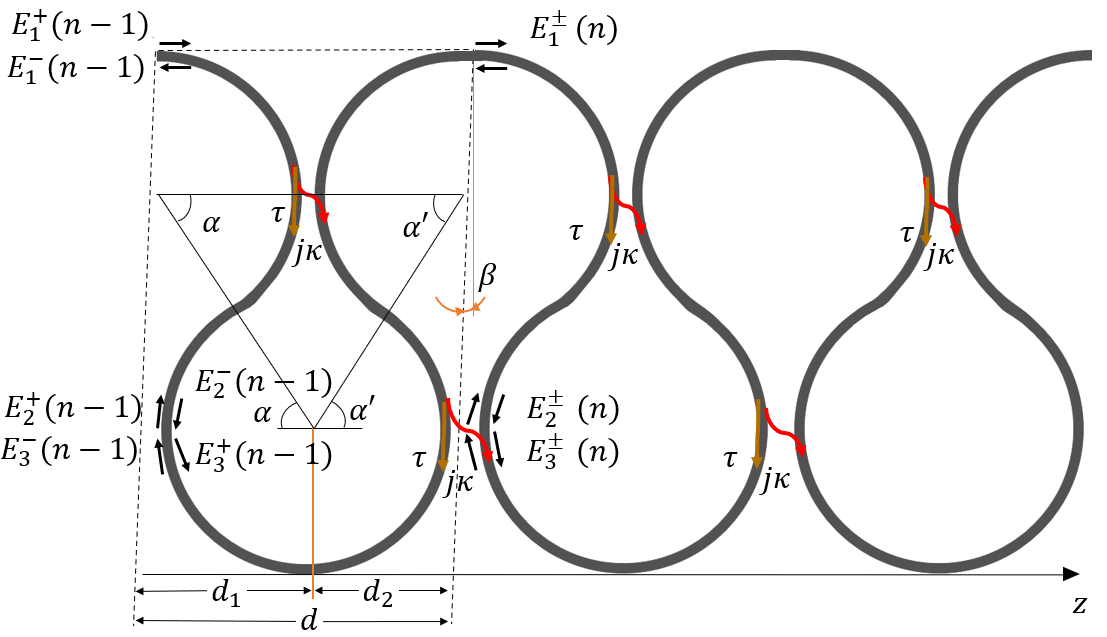}
\caption{Periodic ASOW. The silicon waveguide follows a serpentine path. A unit cell of length $d$ is defined within the two oblique dashed lines. These lines are defined from the apex of the top loops and at an angle $\beta = \alpha - \alpha'$ from the vertical. The lengths $d_1$ and $d_2$ are defined as the distance (at the bottom of the cell) between the dashed oblique lines and the vertical (non-oblique) orange line that goes from the center to the lowest point of the bottom loop. As $\alpha \neq \alpha'$, then $d_1 \neq d_2$. The field amplitudes are defined to the right of the boundaries of the unit cell, with the sign of $E_i^+$, $E_i^-$ corresponding to the sign of the projection of the direction of propagation of that wave with the $z$-axis in the vicinity of the $i$-th port.}
\label{fig:TiltedStructure}
\end{figure}

One fundamental feature of the SIP-related frozen mode is that it corresponds to a particular third order EPD, where three Bloch eigenmodes, one propagating and two evanescent, coalesce at the SIP frequency. For this to happen, all three Bloch eigenmodes collapsing on each other at the EPD must belong to the same one-dimensional irreducible representation of the symmetry group $G_k$ of the Bloch wavevector $k$ {\cite{knox_symmetry_1964}}. This requirement is quite different from the condition for the common symmetry related degeneracy, where the degenerate eigenmodes must belong to the same multidimensional irreducible representation of $G_k$. Since at any given frequency we have just a limited number of Bloch eigenmodes, the easiest way to automatically satisfy the above condition for the SIP existence is to have the symmetry of the waveguide as low as possible. We will apply this guiding principle when choosing the waveguide geometry. 
In a reciprocal periodic waveguide, there will be a pair of reciprocal SIPs with equal and opposite Bloch wavenumbers $k$. Therefore, the existence of EPD in a reciprocal waveguide requires at least six Bloch eigenmodes with the same symmetry – three coalescing Bloch eigenmodes in either direction. Here, we consider a specific example of asymmetric serpentine optical waveguide (ASOW) by applying symmetry-breaking distortion to the symmetric optical waveguide (SOW) in \cite{scheuer_serpentine_2011}.

This paper is organized as follows: In Section \ref{ch:DescriptionWaveguide}, we describe the ASOW. In Section \ref{ch:TransferMatrixFormalism}, we develop a transfer matrix formalism that facilitates obtaining the ASOW eigenmodes. In Section \ref{ch:Stationary Inflection Point} we study the conditions for SIP existence. In Section \ref{ch:TransferFunction} we analyze the scattering problem for a finite ASOW supporting a pair of reciprocal SIPs. In Section \ref{ch:Discussion} we summarize the results.

\begin{figure}[H]
\centering
\includegraphics[width = 0.49\textwidth]{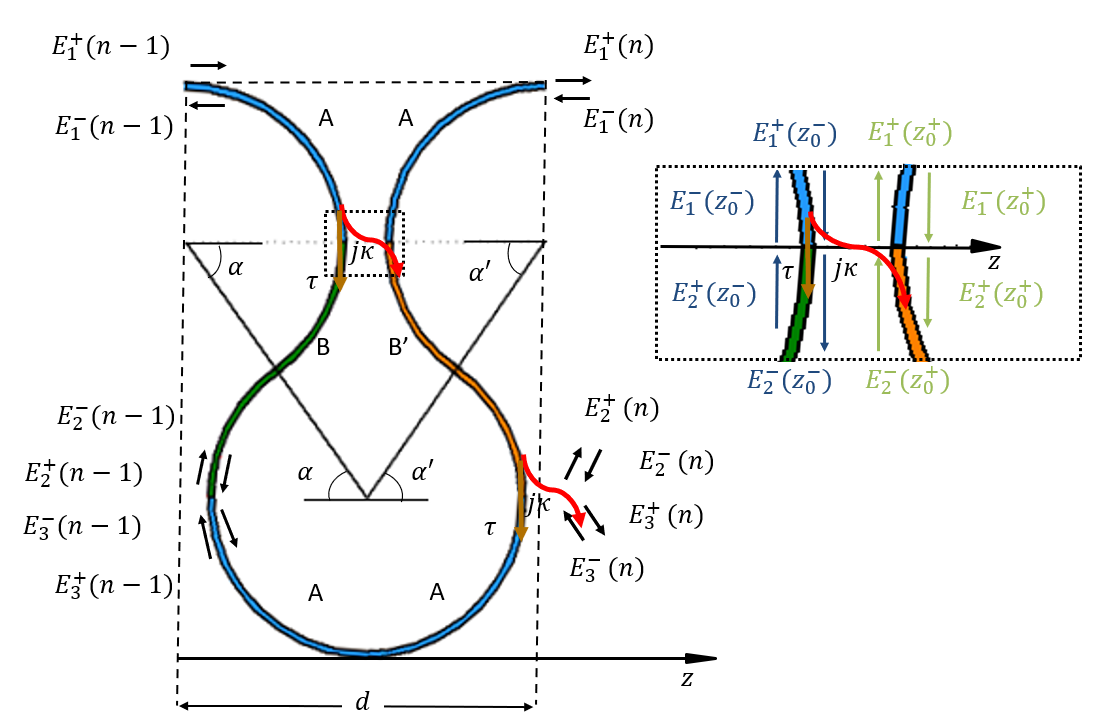}
\caption{The $n$-th unit cell of the ASOW, with its boundaries represented by the two oblique dashed lines from the apex top of the upper loops at an angle $\beta = \alpha - \alpha'$ (the segment A of each top loop in the unit cell is exactly one quarter of a circle). The unit cell waveguide is formed by three different segments: A, B, B'. Segments A are quarters of a loop, marked in blue; Segment B is the waveguide that connects the upper loop with the bottom loop on the left, marked in green, and its length depends on the angle $\alpha$; Segment B' is the waveguide that connects the bottom loop with the upper one, on the right, marked in orange and its length depends on $\alpha'$. The dashed region on the top encloses the lumped, lossless, coupling point $z_0$, which represents the point where the adjacent loops are the closest. Coupling exists also between the bottom loop and the two adjacent unit cells on the left and right (not depicted here).}
\label{fig:TiltedCoupling}
\end{figure}

\section{Geometry of the Asymmetric Serpentine Optical Waveguide}
\label{ch:DescriptionWaveguide}

A SOW related to the one shown in this paper was analyzed in \cite{scheuer_serpentine_2011}. It was shown that that structure supported slow-light mode at regular band edges (RBE), where the group velocity vanishes. Instead, here we focus on a modification of that SOW structure, where the applied deformation and the lack of symmetry in each unit cell enables the occurrence of an SIP. As traditionally assumed \cite{yariv_photonics_2007, vahala_optical_2004} and as in \cite{scheuer_serpentine_2011}, we define the coupling as point-like and lossless, i.e.,
\begin{equation}
    \kappa^2 + \tau^2 = 1,
    \label{eq:Coupler}
\end{equation}

where $\kappa$ and $\tau$ represent the field coupling and transmission coefficients respectively. Both coefficients are constrained to $\kappa, \tau \in [0,1]$.

The ASOW shown in Fig. \ref{fig:TiltedStructure} is a lossless periodic structure in which the adjacent loops are coupled to one another, allowing the formation of resonating optical paths. The waveguide in each unit cell is divided into three segments: A, B \& B'. Segment A is a quarter of a circle with radius $R$. In every unit-cell there are two A segments at the top part, marked blue in Figure \ref{fig:TiltedCoupling}, and additional two A segments that form a half of a circle at the bottom (also marked blue). Segment B is the left-side waveguide connecting the upper and bottom loops and depends directly on the radius $R$ and $\alpha$, marked in green. Segment B' on the right side of the unit cell is similar to segment B but it differs in that it depends on $\alpha'$, as shown in orange in Figure \ref{fig:TiltedCoupling}. The local slope at the transition between to top and bottom loops in Fig. 2 is continuous because the intersection is between two arcs with the same radius $R$ interconnecting at the same angle, either  $\alpha$ or $\alpha'$, therefore there is no slope discontinuity. The phase accumulation associated with each segment is given by

\begin{equation}
\begin{gathered}
    \phi_{a} = k_0 n_w \pi R / 2 ,\\
    \phi_{b} = k_0 n_w 2 \alpha R,\\
    \phi_{b}' = k_0 n_w 2 \alpha' R ,
    \label{PhaseAccumulation}
\end{gathered}
\end{equation}

where $k_0=\omega/c$ is the wavenumber in vacuum, $\omega$ is the angular frequency, $c$ is the speed of light in vacuum, $R$ is the radius of the loops, and $n_w$ is the effective refractive index of the waveguide's mode. $\alpha$ is the angle between the line that crosses the center of the top left and bottom loops and the horizontal axis. $\alpha'$ is the angle between the line that connects the centers of the bottom and top right loops and the horizontal axis. In this ideal design concept, we ignore the gaps between the waveguides in adjacent loops (gaps of the order of $50-100$ nm) on the basis that they are significantly smaller than the radius of the loops (of the order of $10$ $\upmu \text{m}$). The precise gap size is decided based on the design of a realistic coupler, however in this paper for simplicity each coupler is considered as "point-like", satisfying Eq. ({\ref {eq:Coupler}}). As such, the length of the unit cell is given by the diameter of the loops of the ASOW, $d = 2R$.

The key modification of the ASOW in this paper with respect to the SOW in \cite{scheuer_serpentine_2011} is the difference between  $\phi_b$ and $\phi_{b}'$, which breaks the left-right (i.e., longitudinal) symmetry of the unit cell in terms of effective propagation length (akin to the misaligned anisotropic layers studied in \cite{figotin_slow_2011}) and enables the formation of an SIP. The broken symmetry can be understood as a shear deformation since it is realized by imposing $\alpha \neq \alpha'$.  The angle difference $\beta = \alpha - \alpha'$ assumed in this paper to obtain an SIP is very small, so the difference in the lengths of segment B and B' is barely noticeable in Figure \ref{fig:TiltedCoupling}. In Figure \ref{fig:TiltedStructure}, $\beta$ is the angle between the dashed oblique line that defines the boundary at the right side of the unit cell and the vertical orange line crossing the center of the bottom loop and its lowest point. 

\section{Transfer Matrix Formalism }
\label{ch:TransferMatrixFormalism}
We model the electromagnetic guided fields in terms of forward $E_i^{+}$ and backward waves $E_i^{-}$, with $i=1,2,3$; where the superscripts denote the sign of the projection of the direction of propagation of the wave on the $z$-axis. The time convention $e^{j\omega t}$ is implicitly assumed.

The unit cell has six ports, with $E_1^{+}$, $E_2^{+}$ and $E_3^{+}$ propagating towards the right and $E_1^{-}$, $E_2^{-}$ and $E_3^{-}$ propagating towards the left (at or in the vicinity of the ports). The fields are defined at the right of the boundaries. We assume the coupling between adjacent loops to be lumped and lossless. The scattering matrix relating the incoming and outcoming fields at the coupling point $z_0$ is defined in Appendix A and shown in Figure \ref{fig:TiltedCoupling}.

We define a state vector with all the six electric field wave amplitudes as

\begin{equation}
    \boldsymbol{\psi}(n)=\left(\begin{array}{cccccc}
    E_{1}^{+}, \hspace{0.2cm}   E_{1}^{-}, \hspace{0.2cm}  E_{2}^{+}, \hspace{0.2cm}  E_{2}^{-}, \hspace{0.2cm}  E_{3}^{+}, \hspace{0.2cm}  E_{3}^{-}\end{array}\right)^T
    \label{eq:StateVector}
\end{equation}

where $n$ denotes the unit cell number, as seen in Figures \ref{fig:TiltedStructure} and \ref{fig:TiltedCoupling}. The six field terms are calculated at the right side of the same cell, and $T$ denotes the transpose operator. Note that this definition has terms arranged differently from those used in \cite{nada_theory_2017}, and it is the same used in  \cite{scheuer_serpentine_2011}, albeit with a different notation. The state vector on the right side of the $n$-th unit cell is $\boldsymbol{\psi}(n)$. Its "evolution" along the periodic ASOW is described by 

\begin{equation}
    \boldsymbol{\psi}(n)=\underline{\mathbf{T}}_u \boldsymbol{\psi}(n-1)
\label{eq:TransferMatrixUnitCell}
\end{equation}

where $\underline{\mathbf{T}}_u$ is the 6x6 transfer matrix of the unit cell of the ASOW. As the ASOW is reciprocal, the determinant of the transfer matrix satisfies det($\underline{\mathbf{T}}_u)=1$, which causes the eigenvalues of this matrix to come in three reciprocal pairs. This causes the dispersion diagram to show the symmetry that if $k(\omega)$ is a solution of (\ref{eq:Determinant}), then also $-k(\omega)$ is. Hence, the dispersion diagram is symmetric with respect to the center of the Brillouin Zone (BZ), defined here with $\text{Re}(k)$ from $-\pi/d$ to $\pi/d$. The transfer matrix of the unit cell is given by%

\begin{widetext}
\begin{equation}
\resizebox{0.85\textwidth}{!}{$
\ensuremath{\underline{\mathbf{T}}_u}=\left(\begin{array}{cccccc}
0 & -j\frac{\tau}{\kappa} & j\frac{e^{j(\phi_{a}+\phi_{b})}}{\kappa} & 0 & 0 & 0\\
j\frac{\tau}{\text{\ensuremath{\kappa}}} & 0 & 0 & -j\frac{e^{-j(\phi_{a}+\phi_{b})}}{\kappa} & 0 & 0\\
0 & -\frac{\tau e^{-j(\phi_{a}+\phi_{b}')}}{\kappa^{2}} & \frac{\tau^{2}e^{j(\phi_{b}-\phi_{b}')}}{\kappa^{2}} & 0 & j\frac{e^{j2\phi_{a}}}{\kappa} & 0\\
-\frac{\tau e^{j(\phi_{a}+\phi_{b}')}}{\kappa^{2}} & 0 & 0 & \frac{\tau^{2}e^{-j(\phi_{b}-\phi_{b}')}}{\kappa^{2}} & 0 & -j\frac{e^{-j2\phi_{a}}}{\kappa}\\
-\frac{e^{-j(\phi_{a}+\phi_{b}')}}{\kappa^{2}} & 0 & 0 & \frac{\tau e^{-j(\phi_{b}-\phi_{b}')}}{\kappa^{2}} & 0 & -j\frac{\tau e^{-j2\phi_{a}}}{\kappa}\\
0 & -\frac{e^{-j(\phi_{a}+\phi_{b}')}}{\kappa^{2}} & \frac{\tau e^{j(\phi_{b}-\phi_{b}')}}{\kappa^{2}} & 0 & j\frac{\tau e^{j2\phi_{a}}}{\kappa} & 0
\end{array}\right).$}
\label{eq:TransferMatrix}
\end{equation}
\end{widetext}

Its calculation is shown in Appendix A.  Note that if $\phi_b = \phi_{b}'$, this transfer matrix reduces to that of the SOW in \cite{scheuer_serpentine_2011}, where the lossless coupling relation shown in Eq. (\ref{eq:Coupler}) was defined in units of power instead of units of field amplitude used in this paper.

From the Bloch theorem \cite{fujita_bloch_2007}, which states that the field at each unit cell is determined by the field at the adjacent one and a unit cell phase shift, we obtain

\begin{equation}
    \boldsymbol{\psi}(n)=e^{-jkd}\boldsymbol{\psi}(n-1)
    \label{eq:BlochTheorem}
\end{equation}

where $k$ is the Bloch wavenumber of a guided eigenmode and $d$ is the length of the unit cell. By using (\ref{eq:TransferMatrixUnitCell}) and (\ref{eq:BlochTheorem}), we write the eigenvalue problem

\begin{equation}
\ensuremath{\underline{\mathbf{T}}_u \boldsymbol{\psi}(n-1)=\zeta \boldsymbol{\psi}(n-1)}
\label{eq:EigenvalueProblem}
\end{equation}

where $\zeta =e^{-jkd}$. Solving it gives us the eigenvalues and the eigenvectors of the system. When three of these eigenmodes coalesce to a degenerate one with $\Re(k) \neq 0$, the SIP is formed, which is an EPD of order three. The eigenvalue solutions are found from the characteristic equation,

\begin{equation}
\ensuremath{D(k,\omega) \equiv \textrm{det}\left(\underline{\mathbf{T}}_u-\zeta \underline{\mathbf{I}}\right)=0}
\label{eq:Determinant}
\end{equation}

After some algebraic manipulation, we arrive at the following characteristic polynomial 

\begin{equation}
    \begin{split}
    D(k,\omega) &= \zeta^6 - \zeta^5\left(2\frac{\tau^{2}}{\kappa^{2}}cos(\phi_{b}-\phi_{b}')\right) \\
    &+ \zeta^4\left(-2\frac{\tau^2}{\kappa^2} + \frac{\tau^4}{\kappa^4} \right) \\
    &-\zeta^3 \left(\frac{2\cos(4\phi_a+\phi_b+\phi_b')}{\kappa^4} \right)\\   
    &-\zeta^3 \left(\frac{4(\tau^2-\tau^4)\cos(\phi_b-\phi_b')}{\kappa^4}\right) \\
    &-\zeta^2\left(-\frac{\tau^4}{\kappa^4} + 2\frac{\tau^6-2\tau^4-\tau^2}{\kappa^6}\right) \\
    &-2\zeta\frac{\tau^6-2\tau^4 + \tau^2}{\kappa^6} \\
    &+ \frac{\tau^8-4\tau^6+6\tau^4-4\tau^2+1}{\kappa^8}
\label{eq:DispersionRelation}
\end{split}
\end{equation}

The difference between $\phi_b$ and $\phi_{b}'$ is manifested only inside the cosine function, which is an even function. As such, interchanging the values of $\phi_b$ and $\phi_{b}'$ does not change the spectral properties of the ASOW. Notice that due to reciprocity of the ASOW, the solutions come in reciprocal pairs: $k_1$ \& $-k_1$, $k_2$ \& $-k_2$, and $k_3$ \& $-k_3$. In other words, if $\zeta$ is an eigenvalue, $1/\zeta$ is an eigenvalue as well. In the following we represent the wavenumbers in the first BZ, defined here with its center at $\text{Re}(k) = 0$. Because of periodicity, a solution $-k_i$ has Floquet harmonics of the form $-k_i+2\pi m/d$, where $m$ is any integer number. The transfer matrix of the unit cell is similar to the diagonal matrix, 

\begin{equation}
    \underline{\mathbf{T}}_u = \underline{\mathbf{V}} \ \underline{\mathbf{\Lambda}} \ \underline{\mathbf{V}}^{-1}
    \label{eq:DiagonalizedTransferMatrix}
\end{equation}

where $\underline{\mathbf{V}}=[\boldsymbol{\psi}_1|\boldsymbol{\psi}_2|\boldsymbol{\psi}_3|\boldsymbol{\psi}_4|\boldsymbol{\psi}_5|\boldsymbol{\psi}_6]$ is the similarity matrix transformation with  eigenvectors $\boldsymbol{\psi}_i$ as columns, and

\begin{equation}
\underline{\mathbf{\Lambda}}=\left(\begin{array}{cccccc}
e^{-jk_{1}d} & 0 & 0 & 0 & 0 & 0\\
0 & e^{-jk_{2}d} & 0 & 0 & 0 & 0\\
0 & 0 & e^{-jk_{3}d} & 0 & 0 & 0\\
0 & 0 & 0 & e^{jk_{1}d} & 0 & 0\\
0 & 0 & 0 & 0 & e^{jk_{2}d} & 0\\
0 & 0 & 0 & 0 & 0 & e^{jk_{3}d}
\end{array}\right)
\label{eq:ZetaInTermsOfK}
\end{equation}
is the diagonal matrix with the eigenvalues $\zeta_i$, with $i=1,...,6$.  

\section{Exceptional Points of Degeneracy}
\label{ch:Stationary Inflection Point}

EPDs are defined as the points where the eigenmode orthogonality collapses, which means that the algebraic multiplicity of an eigenvalue (the number of identical roots of the characteristic polynomial) is higher than the geometric multiplicity (the number of independent eigenvectors associated with that eigenvalue). This dissonance causes the matrix to not be diagonalizable and it is similar to a matrix containing at least a nontrivial Jordan block. The number of coalesced eigenvectors gives the order of the EPD, with the SIP being an EPD of third order. 

Given the reciprocity of the ASOW, which is seen as a three-way waveguide (analogously to those in \cite{nada_frozen_2021, nada_theory_2017, paul_frozen_2021, apaydin_experimental_2012}), this waveguide supports at any given frequency three pairs of reciprocal Bloch eigenmodes, which allow only degeneracies of order 2, 3, 4 and 6 to form. For the ASOW to exhibit an SIP at a generic point, that is, away from the center or the boundaries of the BZ, all three Bloch eigenmodes with the same sign of $\text{Re}(k)$ should coalesce. In the case of the undistorted SOW, the symmetry of a generic point of the BZ has a single nontrivial operation - the glide mirror plane normal to the $x$-direction. Any Bloch eigenmode, propagating or evanescent, of the undistorted structure is either even or odd with respect to the above symmetry operation. Normally, two of the three eigenmodes have the same parity, while the third one has the opposite parity. The states with the opposite parity do not usually coalesce and, thus, are less likely to participate in SIP formation. On the other hand, the two eigenmodes with the same parity can coalesce and form a regular band edge (RBE) \cite{scheuer_serpentine_2011}. To facilitate the coalescence of all three eigenmodes of $\text{Re}(k)$ with the same sign, we break the glide plane symmetry by applying the shear distortion described in Figure \ref{fig:TiltedStructure} on the undistorted SOW.

In this paper we focus on finding SIPs, which are found as inflection points at ($k_s,\omega_s$) in the dispersion diagram, locally approximated as

\begin{equation}
    \omega-\omega_s \propto \left(k-k_s\right)^3.
    \label{eq:SIP_approx}
\end{equation}

The existence of an SIP indicates that the structure (ASOW in our case) possesses a frozen mode regime, exhibiting huge diverging amplitudes and low group velocity \cite{figotin_slow_2011}. At frequencies in the vicinity of the SIP, the guided field is a superposition of a propagating and two evanescent Bloch modes, which develop a strong singularity close to the SIP frequency while remaining nearly equal and opposite in sign at the boundary of the structure, satisfying boundary conditions.

The advantage of the SIP compared to EPDs of even order such as RBE and degenerate band edge, or DBE (order $2$ and $4$, respectively), is that it can exhibit a good coupling efficiency \cite{gutman_stationary_2011}, with a significant fraction of the incident light coupling into the waveguide. The high coupling efficiency  allows SIP-exhibiting structures to interact effectively with external devices. This is in contrast to structures exhibiting RBEs and DBEs where the impedance mismatch is substantially larger \cite{othman_theory_2016}.

An SIP is defined as a third-order EPD, which means that (\ref{eq:DiagonalizedTransferMatrix}) does not hold anymore, and that $\underline{\mathbf{T}}_u$ is degenerate with two reciprocal eigenvalues of algebraic multiplicity 3 and geometric multiplicity 1 (i.e., there are only two eigenvalues $\zeta_s=e^{-jk_sd}$ and $\zeta_s^{-1}=e^{jk_sd}$, repeated three times each, and two eigenvectors associated to those eigenvalues).

\subsection{Analytic Dispersion Relation for an SIP}

We derive analytically the system of equations that constraints the values of the ASOW parameters $\kappa$, $R$, $\alpha$ and $\alpha'$ such that the ASOW exhibits an SIP. At the SIP angular frequency $\omega_s$ the characteristic equation of the system, found in (\ref{eq:Determinant}), can be cast in a simple way because it has two degenerate Floquet-Bloch eigenwaves. Hence, the characteristic equation evaluated at $\omega_s$ must have the form

\begin{equation}
\ensuremath{D\left(k,\omega_s\right)=\left(\zeta-\zeta_{s}\right)^{3}\left(\zeta-\zeta_{s}^{-1}\right)^{3}} = 0
\label{eq:SIP_DispRel}
\end{equation}

By equating the coefficients of this polynomial with those of the dispersion relation in Equation (\ref{eq:DispersionRelation}) evaluated at $\omega_s$, we derive the following five necessary conditions:

\begin{equation}
     \begin{split}
    &2\frac{\tau^{2}}{\kappa^{2}}\cos\left(\phi_{b}-\phi_{b}'\right)=3\left(\zeta_{s}+\zeta_{s}^{-1}\right), \\
    &\frac{2\cos\left(4\phi_{a}+\phi_{b}+\phi_{b}'\right)+4\left(\tau^{2}-\tau^{4}\right)\cos\left(\phi_{b}-\phi_{b}'\right)}{\kappa^{4}} =\\
    & \;\;\;\;\;\;\; = \left(\zeta_{s}^{3}+\zeta_{s}^{-3}\right)+ 9\left(\zeta_{s}+\zeta_{s}^{-1}\right), \\
    &2\frac{\left(\tau^{6}-2\tau^{4}+\tau^{2}\right)\cos\left(\phi_{b}-\phi_{b}'\right)}{\kappa^{6}}=3\left(\zeta_{s}+\zeta_{s}^{-1}\right),\\
    &\frac{\tau^{4}}{\kappa^{4}}-2\frac{\tau^{6}-2\tau^{4}+\tau^{2}}{\kappa^{6}}=3\left(\zeta_{s}^{2}+\zeta_{s}^{-2}\right) + 9, \\
    &\frac{\tau^{8}-4\tau^{6}+6\tau^{4}-4\tau^{2}+1}{\kappa^{8}}=1 
\label{eq:IdentitiesSIP}
\end{split}
\end{equation}

Equations (\ref{eq:IdentitiesSIP}) must be satisfied for the ASOW to exhibit an SIP at $\omega_s$. The last equation is automatically verified when the coupling and transmission coefficients satisfy Eq. (\ref{eq:Coupler}), i.e., when each coupling is lossless. The fourth equation does not depend on $\phi_a$, $\phi_b$, and $\phi_{b}'$, and the choice of the coupling and transmission coefficients determines the value of $\zeta_s$, hence the wavenumber $k_s$ of the SIP. The first and the third, are not independent: after equating their right hand sides, we get an equation in $\tau$ and $\kappa$ only, which is verified assuming the coupling and transmission satisfy Eq. (\ref{eq:Coupler}). Therefore, either the first or the third equation is useful to determine the phase difference  $\phi_b - \phi_{b}'$ once the coupling and transmission coefficients have been determined. The second equation is useful to determine the phase term $4\phi_a + \phi_b + \phi_{b}'$, which is the total phase accumulated in a unit cell when we do not consider coupling effects. This shows that there are various combinations of the lengths of the segments A, B and B' that lead to an SIP.

In order to quantify the coalescence of the eigenvectors, we use the concept of "coalescence parameter" introduced in \cite{abdelshafy_exceptional_2019} for the DBE and in \cite{nada_frozen_2021} for the SIP. Here, we use a coalescence parameter $\sigma$ defined similarly to that in \cite{nada_frozen_2021}, as
\begin{equation}
    \begin{split}
      &\sigma = \sqrt{\sigma'^2 + \sigma''^2} 
      \\
        &\sigma' = \sqrt{\sum_{m=1,n=2,n>m}^3 |\theta_{mn}|^2} \\
        &\sigma'' = \sqrt{\sum_{m=4,n=5,n>m}^6 |\theta_{mn}|^2} \\
        &\cos(\theta_{mn}) = \frac{|<\boldsymbol{\psi}_m|\boldsymbol{\psi}_n>|}{\|\boldsymbol{\psi}_m\|\|\boldsymbol{\psi}_n\|}
    \end{split}
    \label{eq:SIP_CoalescenceParameter}
\end{equation}

The coalescence parameter is calculated by organizing the eigenvectors $\boldsymbol{\psi}_i$ in two sets of three vectors, associated with $\zeta_i$, and $1/\zeta_i$, respectively, with $i=1,2,3$. Then, we calculate the euclidean distance of the angles between all the combinations in the set with respect to the origin. 
Here, $\theta_{mn}$ is the angle between two 6-dimensional complex vectors $\boldsymbol{\psi}_m$, $\boldsymbol{\psi}_n$, which is defined as stated in Equation (\ref{eq:SIP_CoalescenceParameter}) using the inner product $<\boldsymbol{\psi}_i | \boldsymbol{\psi}_j> = \boldsymbol{\psi}_i^\dagger{}\boldsymbol{\psi}_j$ with the dagger symbol $\dagger{}$ representing the complex conjugate transpose operation and $\|\boldsymbol{\psi}_m\|$ denotes the norm of $\boldsymbol{\psi}_m$ \cite{nada_frozen_2021}. In this paper, we calculate the coalescence parameter using the norm based on the euclidean distance between the parameters $\theta_{mn}$, $ \forall m,n$, and zero  \cite{axler_calculus_1985}, instead of using the arithmetic average used in \cite{nada_frozen_2021,abdelshafy_exceptional_2019}. The reason for this change is that the optimization algorithm converges faster using the euclidean distance than the arithmetic average, as long as the algorithm does not generate a lot of points far from the optimization goal (known as outliers) \cite{janocha_loss_2017}.
The coalescence parameter is always positive and smaller than 1, with $\sigma = 0$ (the origin) indicating perfect coalescence of each set of three eigenvectors. This point constitutes an SIP.

\subsection{ASOW with SIP}

In this section we show that the proposed ASOW exhibits an SIP through the proper tuning of the various structure parameters. For practical purposes, the SIP wavelength is set at $1550$ nm; the waveguide consists of a silicon over insulator structure, and the Si waveguide is assumed to have a height of $230$ nm and a width of $430$ nm. At this wavelength, the lowest TE-like mode has an effective refractive index of $n_w=2.362$, as can be seen in \cite{scheuer_serpentine_2011}. In \cite{dattner_2011} it is also seen that the variation of the refractive index is negligibly small in the  frequency range of interest. In order for the assumption that the structure is lossless at optical frequencies to be reasonable, we restrict ourselves to ASOWs comprising loops with radius $R \geq 10$ $\upmu \text{m}$  \cite{cherchi_dramatic_2013} to minimize radiation losses.

\begin{widetext}
\begin{figure}[H]
    \centering
    \begin{minipage}{0.8\paperwidth}
    \centering
    \subcaptionbox{}
            {\includegraphics[width=0.43\linewidth]{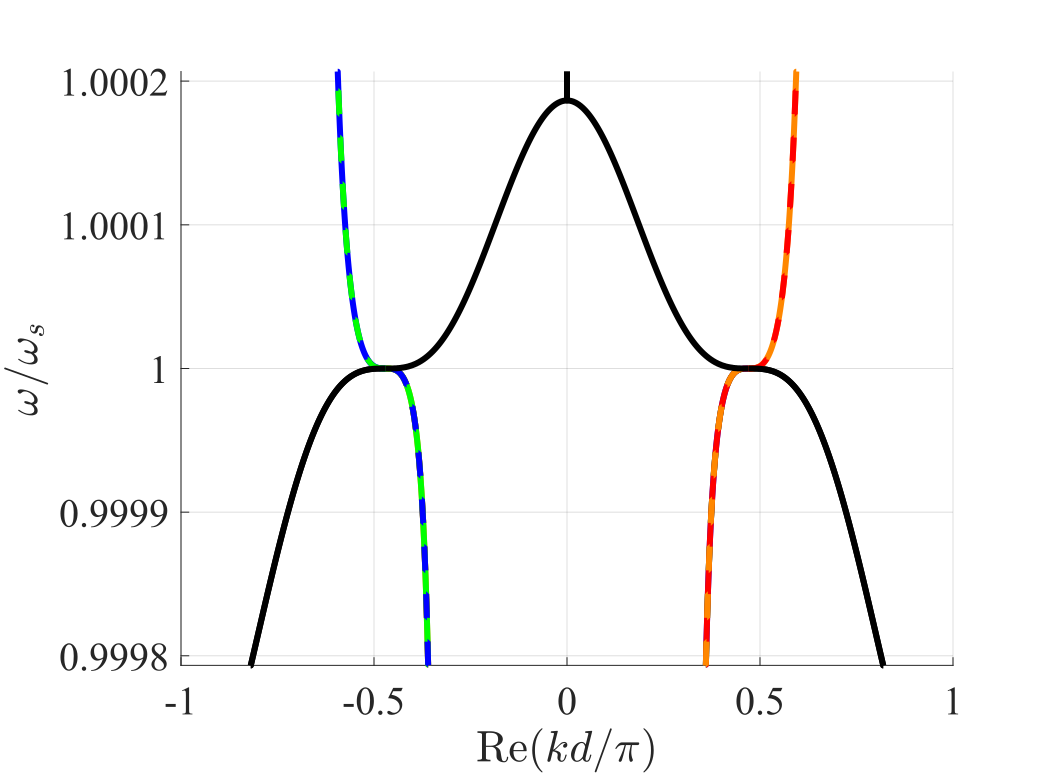}}
    \subcaptionbox{}
    {\includegraphics[width=0.43\linewidth]{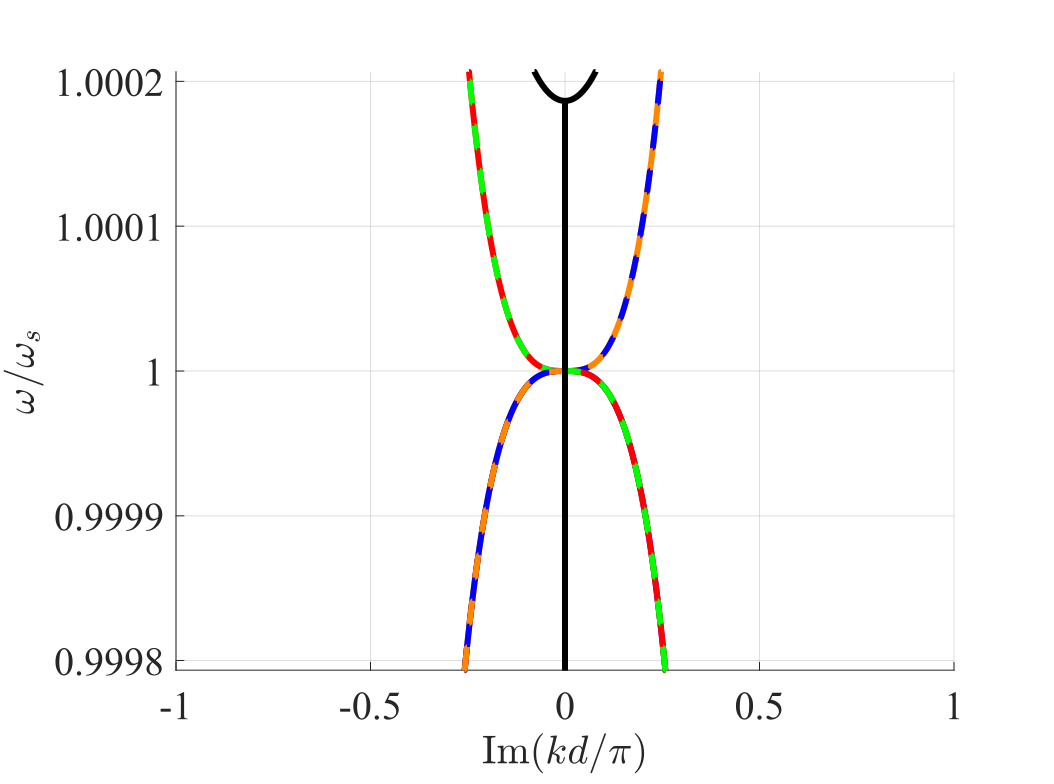}}
    \end{minipage} \quad
    \begin{minipage}{0.8\paperwidth}
    \centering
  \subcaptionbox{}
    {\includegraphics[width=0.43\linewidth]{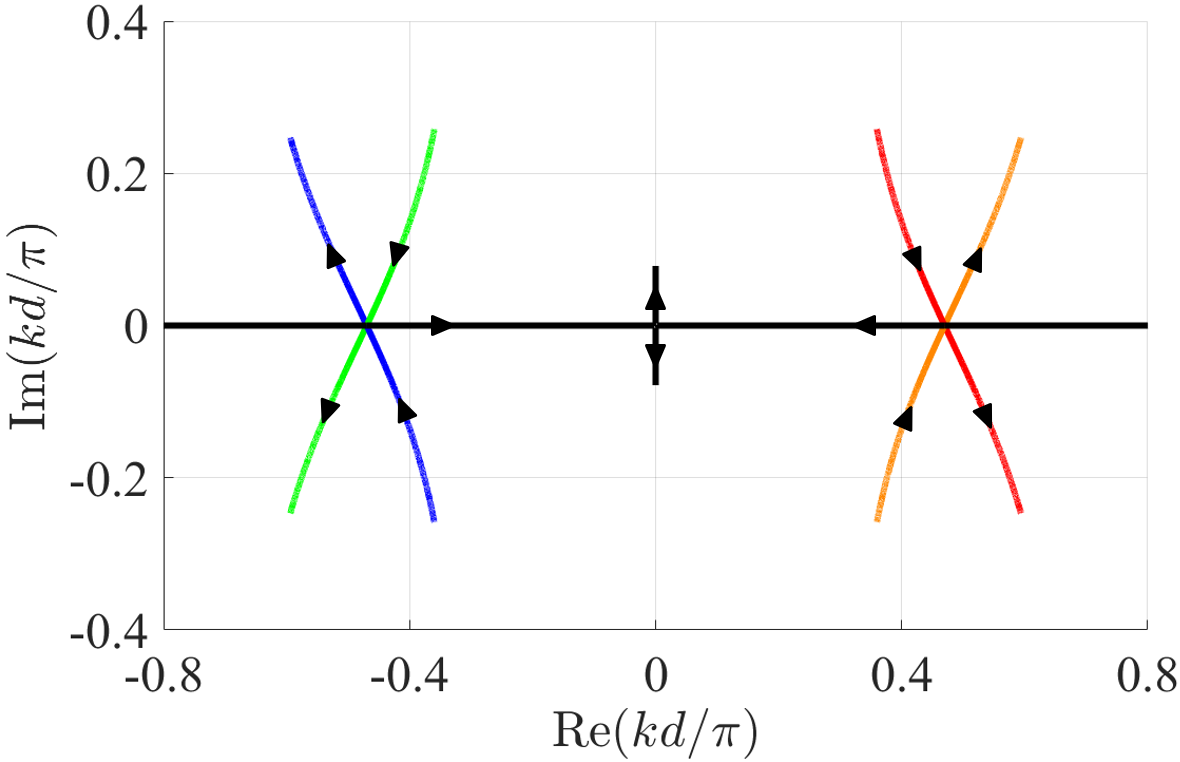}}
    \subcaptionbox{}
    {\includegraphics[width=0.43\linewidth]{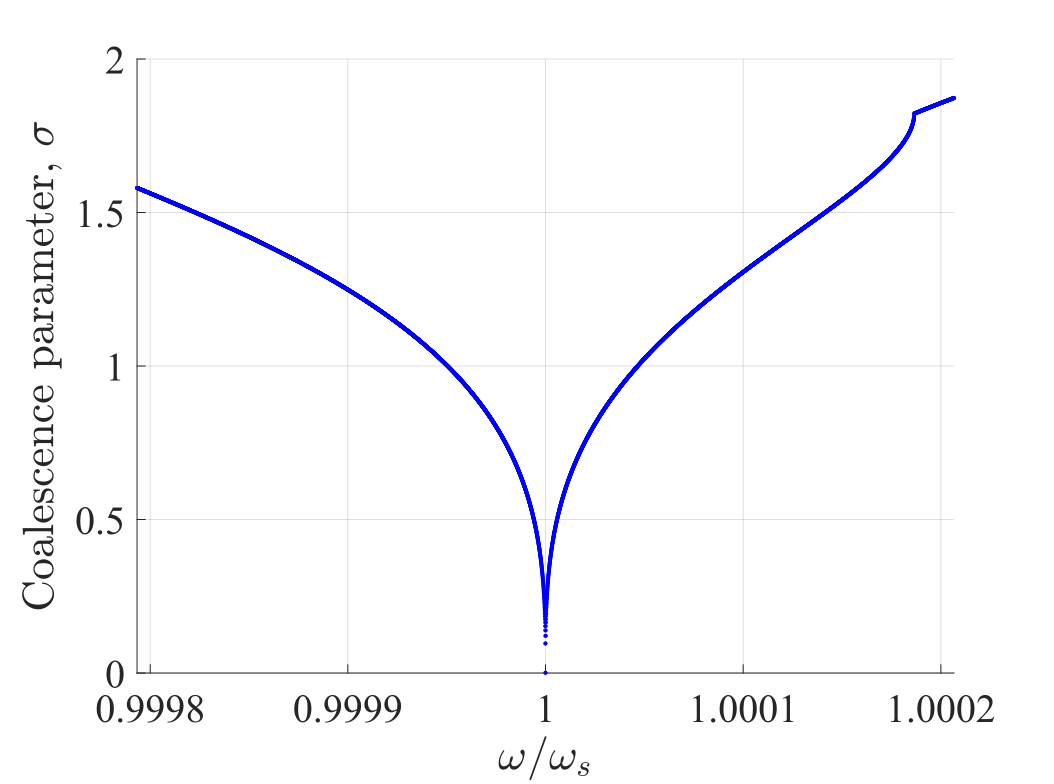}}
    \end{minipage}
    \begin{minipage}{0.8\paperwidth}
    \caption[\linewidth]{Modal dispersion diagram showing the formation of an SIP, using structure parameters: $R = 10$ $\upmu \text{m}$, $\alpha =  66.02 ^{\circ} $, $\alpha' =  56.18 ^{\circ} $ and $\kappa = 0.49$. (a) Real part of the wavenumber versus angular frequency in the fundamental BZ. Solid black: mode with purely real $k$; dashed colors: modes with complex $k$ (overlapping dashed colors imply two overlapping branches).  It is clear that three curves meet at an inflection point, with reciprocal $k$ and $-k$ positions. (b) Imaginary part of $k$ versus angular frequency. At the SIP, $\text{Im}(k)=0$. (c) Alternative representation of the dispersion diagram in the complex $k$ space. The coalescing of the three branches is clearer in this figure, with arrows pointing in the direction of increasing frequency. (d) Coalescence parameter $\sigma$ versus angular frequency, with $\sigma$ vanishing at the SIP frequency.} 
    \end{minipage}
\label{fig:Dispersion-Diagram-of}
\end{figure}
\end{widetext}

Figure 3 depicts the dispersion diagram of the eigenmodes of the ASOW unit cell shown in Figure \ref{fig:TiltedStructure}. It exhibits an SIP that can be seen by the coalescing of the three branches in both the real and imaginary parts.

In addition to the SIP, we also find an RBE not far from the SIP. The distance between the RBE and the SIP most likely decreases with an increasing loop radius. The reasoning behind it is that a larger radius causes the structure to support multiple resonances and reduces its free spectral range, although more work has to be done to investigate how to design RBEs far from the SIP.
The fact that both RBEs and SIPs are both found in a small frequency range could be problematic when realizing lasers. Therefore, learning how to optimize the size of the loops in order to balance between bending (radiation) losses and the formation of RBEs near the SIP frequency is important for exploiting the potential of an SIP. 

\section{Analysis of a finite-length structure}
\label{ch:TransferFunction}

We analyze an ASOW with a finite-length $L=dN$, where $d$ is the period of the unit cell and $N$ is the number of unit cells. This finite-length structure is shown in Fig. \ref{fig:SOW-TransferFunction}. 

\subsection{Field amplitudes along the ASOW}

\begin{figure}[H]
\centering
\includegraphics[width = 0.46\textwidth]{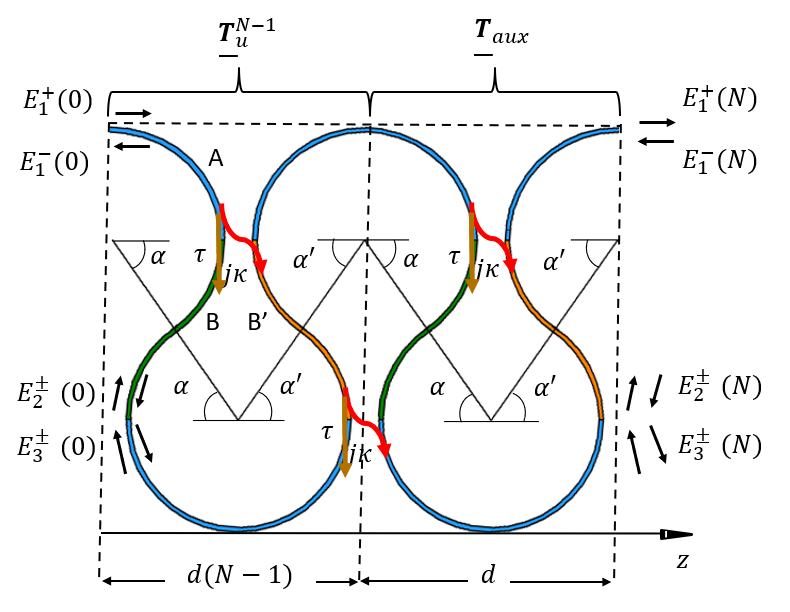}
\caption{Schematic of a finite-length ASOW consisting of $N$ unit cells most of which are described by the transfer matrix $\underline{\mathbf{T}}_u^{N-1}$ and a last unit cell without the second coupling, which connects the ports $2$ and $3$ as defined in Figure \ref{fig:TiltedCoupling}. This last unit cell is described by the transfer matrix $\underline{\mathbf{T}}_{aux}$ and has a length $d$, as we neglect the gap between adjacent loops. All the units cells are defined within parallel oblique dashed lines as described in Section \ref{ch:DescriptionWaveguide}.} 
\label{fig:SOW-TransferFunction}
\end{figure}

The evolution of the field amplitudes from one unit cell to the next is given by Equation (\ref{eq:TransferMatrixUnitCell}). To find the field amplitudes at each unit cell of the structure shown in Fig. \ref{fig:SOW-TransferFunction}, we need to find the field amplitudes (i.e., the state vector) at either end of the structure.

We consider the state vector at the left boundary of the first unit cell of the structure, $\mathbf{\psi_0}=\boldsymbol{\psi}(n=0)$. The state vector at the end of the ASOW made of $N$ cascaded unit cells is given by

\begin{equation}
	\boldsymbol{\psi}(N) = \underline{\mathbf{T}} \mathbf{\psi_0}
    \label{eq:StateVectorFromZtoZero}
\end{equation}

Dividing the ASOW in unit cells as shown in Figure \ref{fig:SOW-TransferFunction}, we find that

\begin{equation}
	\underline{\mathbf{T}} = \underline{\mathbf{T}}_{aux} \ \underline{\mathbf{T}}_u^{N-1} = \underline{\mathbf{T}}_{aux} \ \underline{\mathbf{V}} \ \underline{\mathbf{\Lambda}}^{N-1} \  \underline{\mathbf{V}}^{-1}
\label{eq:TransferMatrixInTermsOfUnitCells}
\end{equation}

where $\underline{\mathbf{T}}_{aux}$ is the transfer matrix of a unit cell without the second coupling point and it is given in Appendix A. 

\begin{figure}[H]
\centering
    \includegraphics[width = 0.48\textwidth]{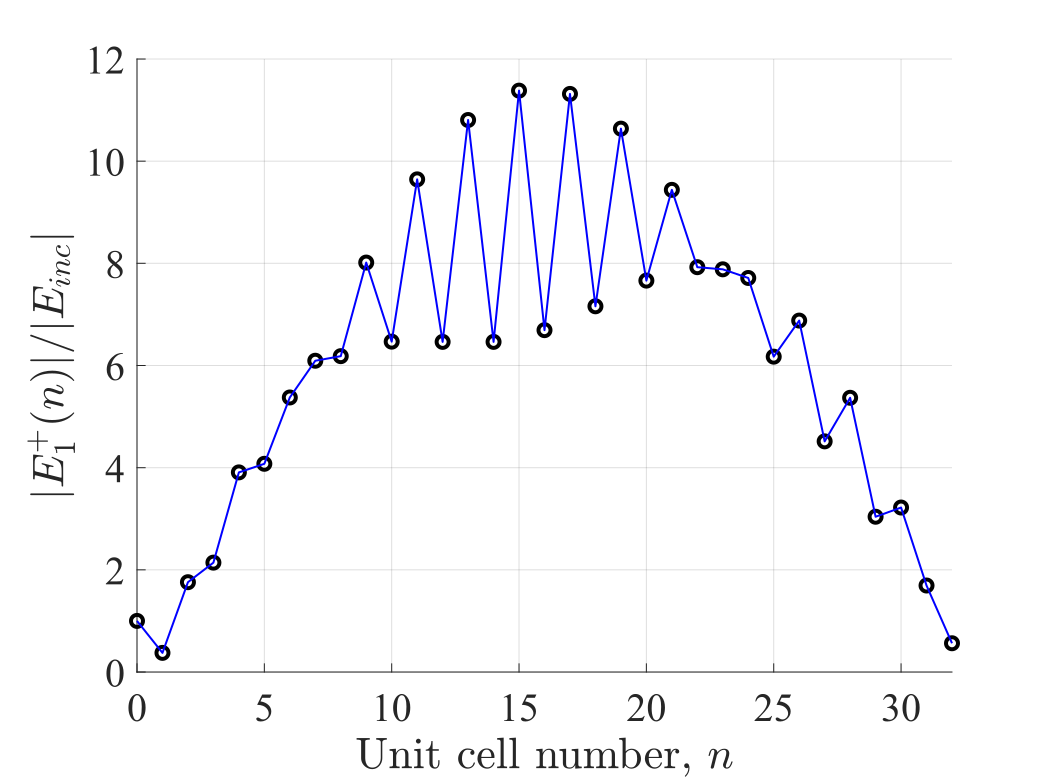}
\hfill
    \includegraphics[width = 0.48\textwidth]{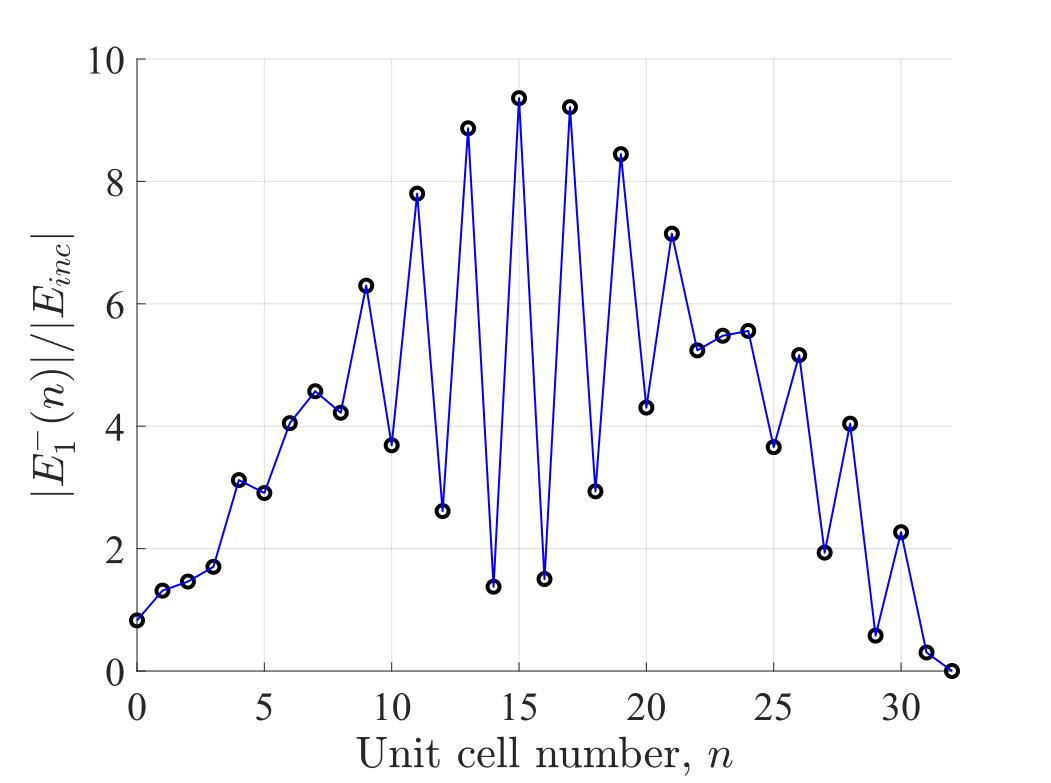}
\hfill
    \includegraphics[width = 0.48\textwidth]{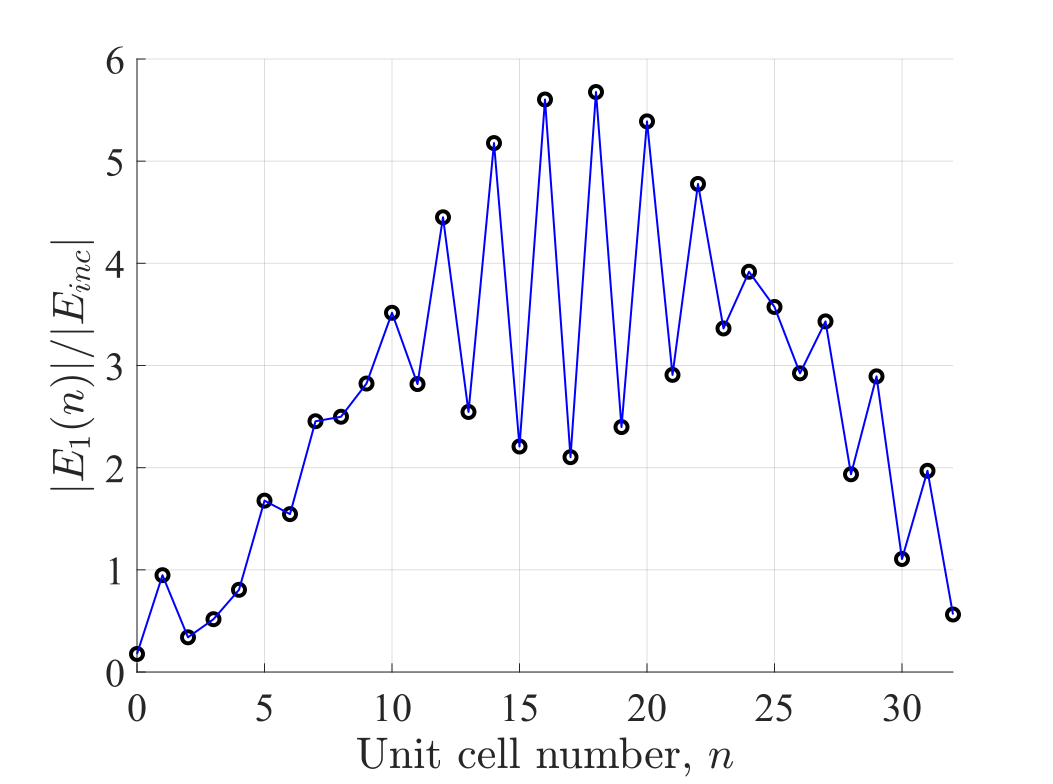}
\caption[\linewidth]{Magnitude of the forward and backward fields throughout the SIP-exhibiting finite-length structure against the position of the unit cell, with $n \in [0,N]$ for $N = 32$ at the SIP frequency $\omega_s$. Notice that the BC from Equation (\ref{eq:BoundCond}) are satisfied. (a) Magnitude of $E_1^+(n)$ for $n \in [0,N]$. (b) Magnitude of $E_1^-(n)$ for $n \in [0,N]$. (c) Magnitude of $E_1(n) = E_1^-(n) + E_1^+(n)$ for $n \in [0,N]$.} 
\label{fig:E1}
\end{figure}

The diagonal matrix $\underline{\mathbf{\Lambda}}$ is defined as in Equation (\ref{eq:DiagonalizedTransferMatrix}).
At an SIP, the transfer matrix is non-diagonalizable and similar to a matrix containing two Jordan blocks. \cite{nada_theory_2017}

\begin{figure}[H]
\centering
    \centering
    \includegraphics[width = 0.4\textwidth]{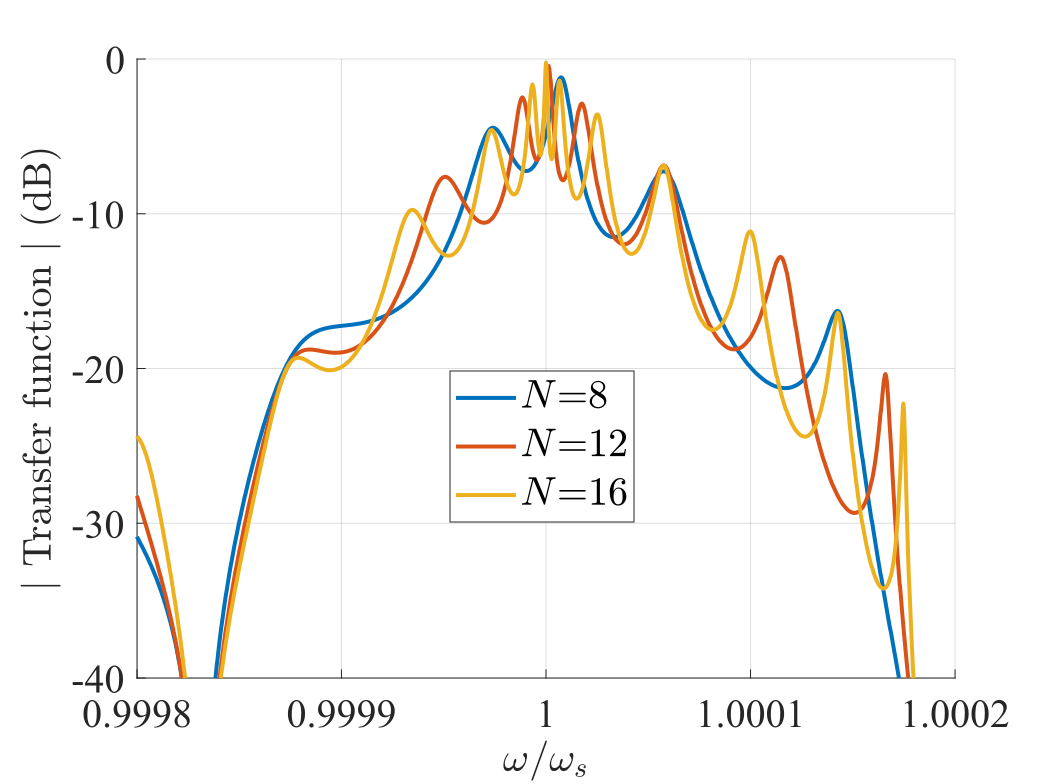}
    \centering
    \includegraphics[width = 0.4\textwidth]{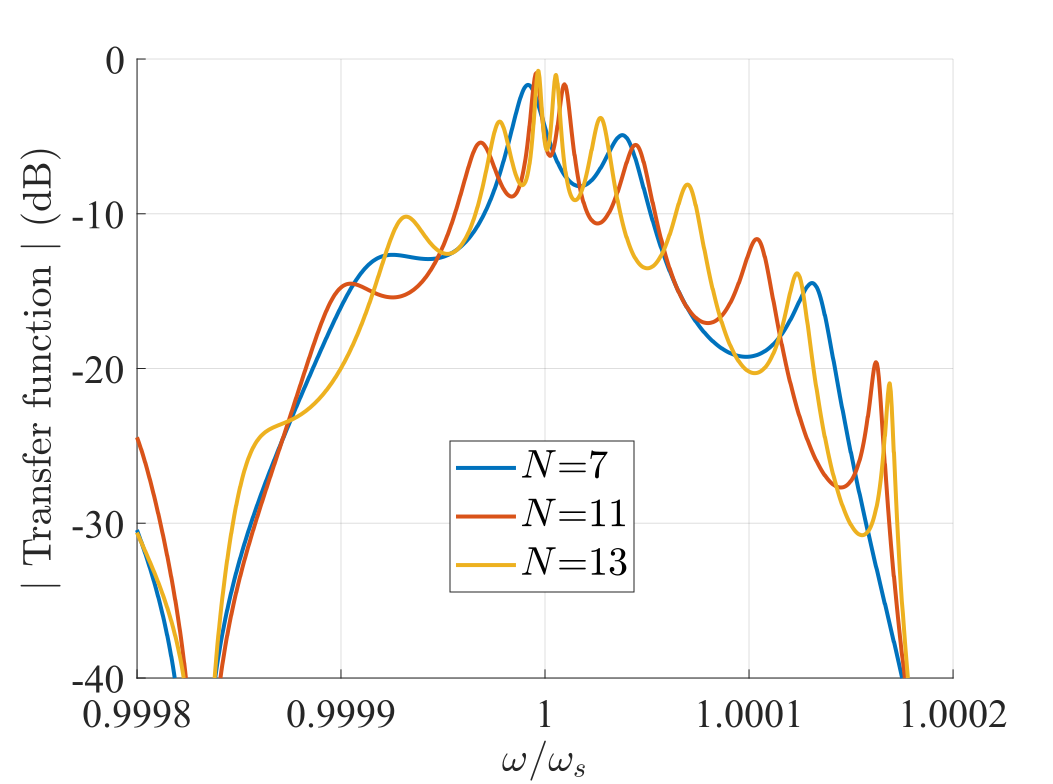}
\caption[\linewidth]{Magnitude of the transfer function (in dB) of the finite-length structure near the SIP frequency for (a) even and (b) odd number of unit cells $N$.} 
\label{fig:Transfer-Function-=00005BdB=00005D}
\end{figure}

We assume the ASOW is excited by an incoming wave $E_{1}^{+}(0) = E_{inc}$ from the left, and the right end is terminated on a dielectric waveguide with the same shape and characteristic impedance of the waveguide used to form the ASOW. Considering the definitions in Fig. \ref{fig:SOW-TransferFunction}, the fields defining the Boundary Conditions (BC) of the waveguide are  

\begin{equation}
    \begin{split}
    &E_{1}^{+}(0)= E_{inc} \\
    &E_{1}^{-}(N)=0 \\
    &E_{2}^{+}(0)=E_{3}^{-}(0) \\
    &E_{3}^{+}(0)=E_{2}^{-}(0)\\
    &E_{2}^{-}(N)=E_{3}^{+}(N) \\
    &E_{3}^{-}(N)=E_{2}^{+}(N) 
    \label{eq:BoundCond}
    \end{split}
\end{equation}

Applying these BC to the state vector "evolution" described in Eq. (\ref{eq:StateVectorFromZtoZero}) gives the field amplitudes at either side of the boundary. By applying Equation (\ref{eq:TransferMatrixUnitCell}), we obtain the field amplitudes at each unit cell from those at $n=0$. The results for $|E_1^+(n)|$, $|E_1^-(n)|$ and  $|E_1(n)|$ over $n \in [0,N]$ are shown in Figure \ref{fig:E1} where $N = 32$. The frozen mode regime, which is characteristic of light traveling with null group velocity followed by a dramatic enhancement of the field amplitudes \cite{figotin_slow_2006}, is in full display. The amplitudes of the fields in the middle of the finite-length structure are substantially larger than those located at its edges, where the BC shown in Equation (\ref{eq:BoundCond}) are satisfied. This frozen mode regime is visible in the magnitude of both the forward and backward waves, as seen in Fig. \ref{fig:E1}, where  $|E_1^+(n)|$, $|E_1^-(n)|$ and their sum $|E_1(n)| = |E_1^-(n) + E_1^+(n)|$ peak around the center of the finite-length waveguide.

\begin{figure}[H]
\centering
    \centering
    \includegraphics[width = 0.47\textwidth]{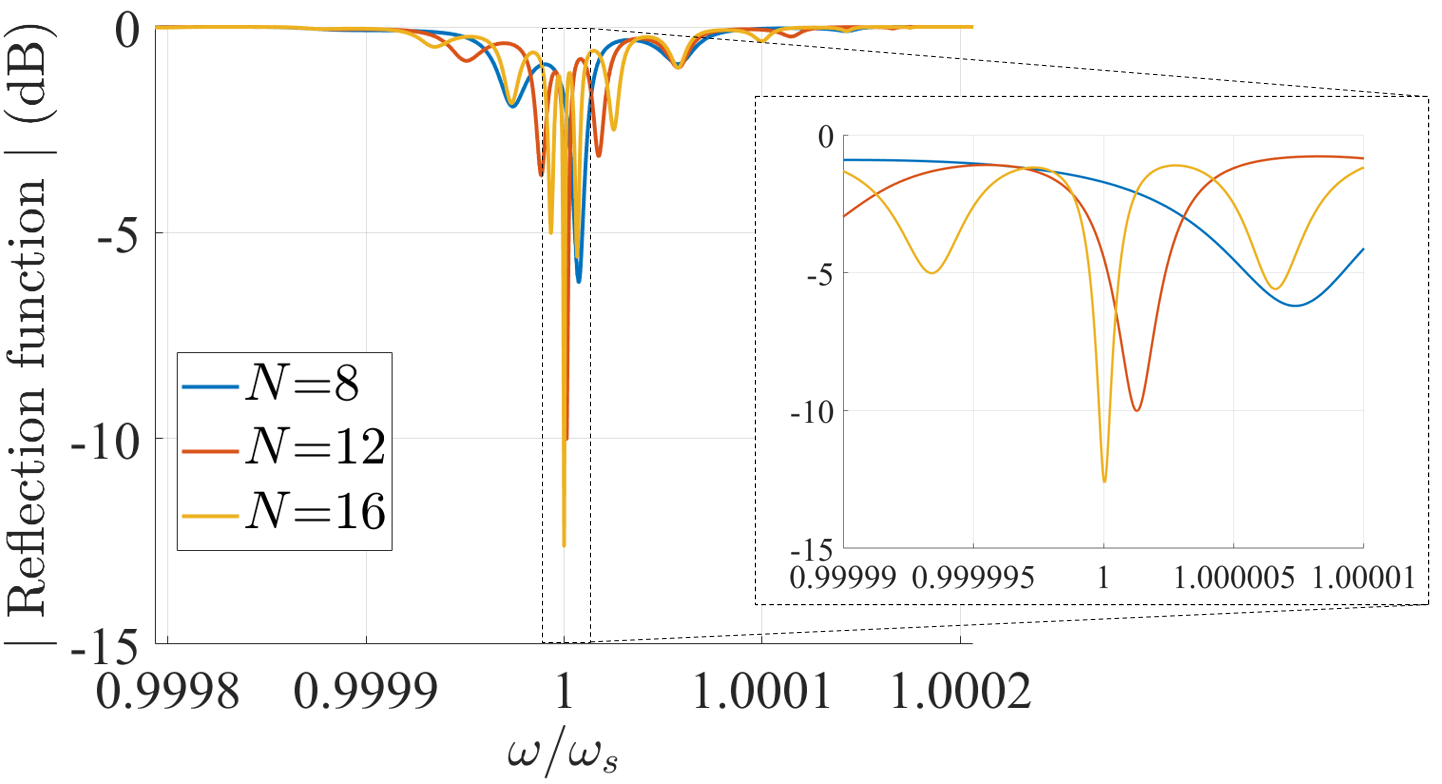}
    \centering
    \includegraphics[width = 0.47\textwidth]{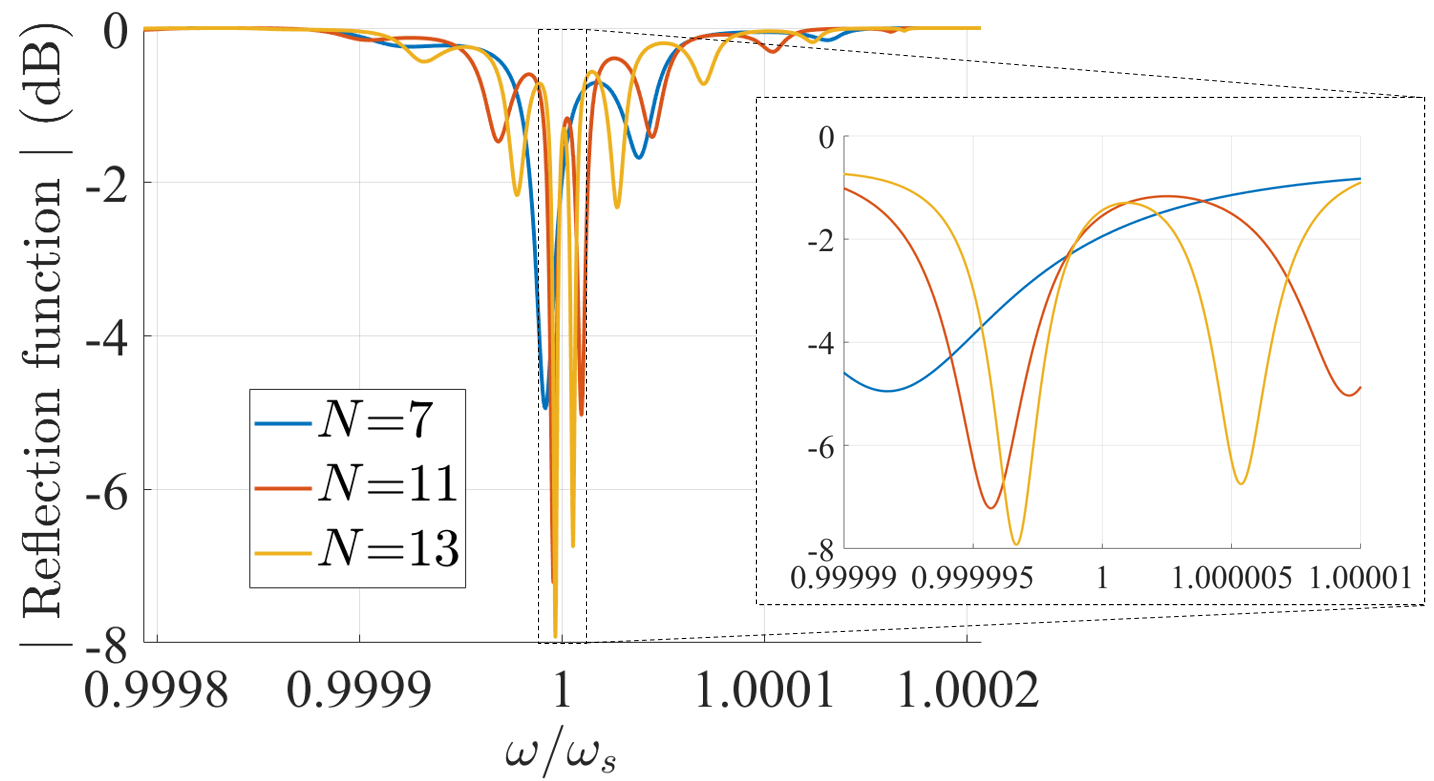}
\caption[\linewidth]{Magnitude of the reflection function (in dB) of the finite-length structure near the SIP frequency for (a) even and (b) odd number of unit cells $N$.} 
\label{fig:Refl-Function}
\end{figure}

\subsection{Transfer function}

Besides the spatial evolution of the field amplitudes along the finite-length structure, we are interested in finding the proportion of light that makes it through the waveguide and the proportion of light that is reflected from it. We define the transfer function 

\begin{equation}
\ensuremath{T_{f}=\frac{E_{out}}{E_{inc}}=\frac{E_{1}^+(N)}{E_{1}^{+}(0)}}\label{eq: Transfer Function}
\end{equation}

as the ratio between the forward field amplitude at the output of the ASOW and the incident one. We also define the reflection function 

\begin{equation}
\ensuremath{R_{f}=\frac{E_{refl}}{E_{inc}}=\frac{E_{1}^-(0)}{E_{1}^{+}(0)}}\label{eq:reflfunction}
\end{equation}

as the ratio between the backward field amplitude at the input of the ASOW and the incident one. The transfer function is equivalent to the s-parameter $S_{21}$ and $R_f$ is equivalent to $S_{11}$.

Figures \ref{fig:Transfer-Function-=00005BdB=00005D} and \ref{fig:Refl-Function} respectively show the magnitude of the transfer and the reflection functions (in dB) of an ASOW comprising $N$ cascaded unit cells, for several values of $N$. The parameters of the structure are chosen to satisfy the conditions from Equation (\ref{eq:IdentitiesSIP}) to exhibit the SIP shown in Figure 3.

The transmission curves reach their maximum in the vicinity of the SIP frequency, where the reflection curves reach their minimal level. The resonance closest to the SIP frequency is denoted hereon as SIP resonance. The distance between peaks in each curve shrinks as $N$ increases. Notice that the peaks in Figs.  \ref{fig:Transfer-Function-=00005BdB=00005D}(a) for even $N$  are more bundled together around the SIP frequency $\omega_s$ than for odd $N$. 

\begin{figure}[H]
\centering
    \includegraphics[width = 0.48\textwidth]{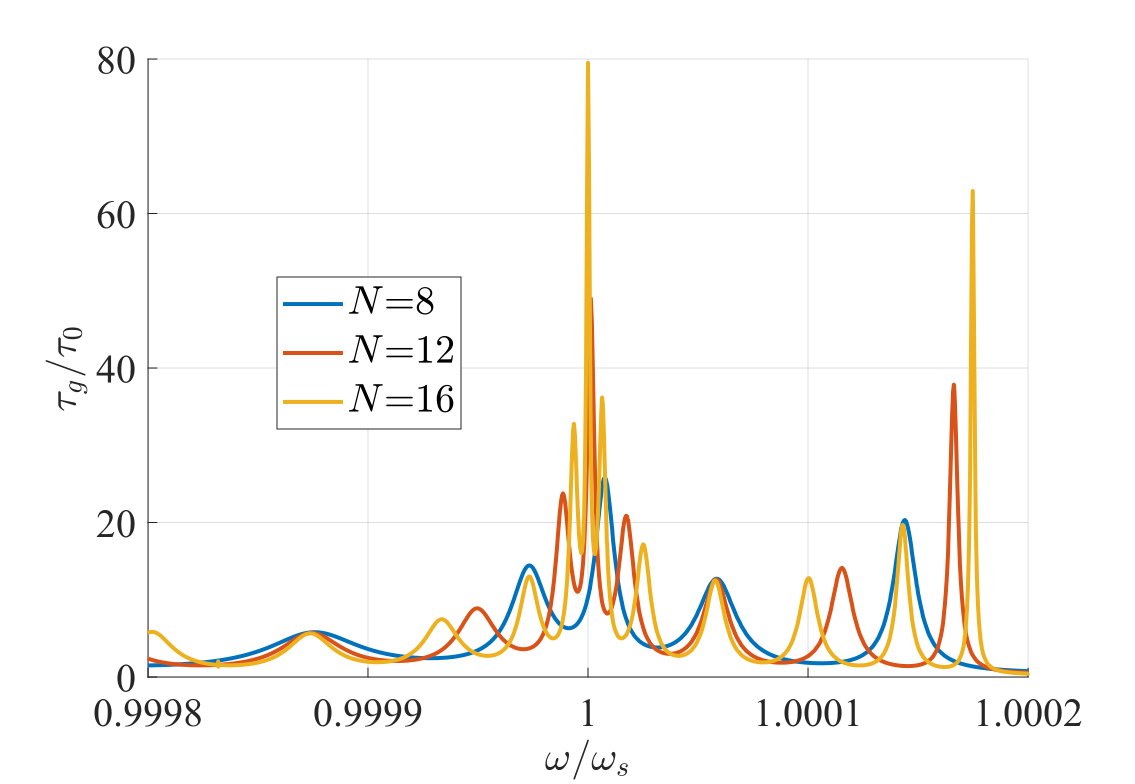}
\hfill
    \centering
    \includegraphics[width = 0.48\textwidth]{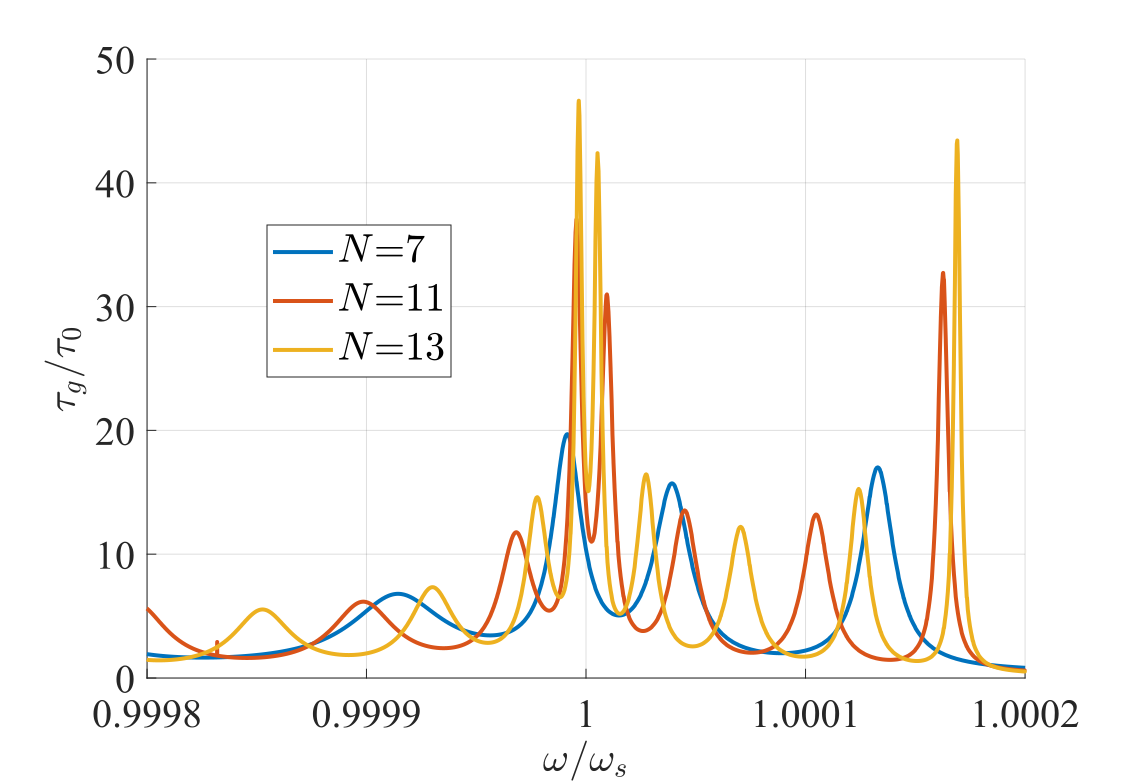}
\caption{Normalized group delay versus normalized frequency for different number of unit cells $N$ for (a) even number of unit cells $N$ and (b) odd number of unit cells.}
\label{fig:Group-Delay-=00005Bs=00005D}
\end{figure}

\subsection{Group delay and quality factor}

The quality factor ($Q$) of a cavity is a measure of the energy lost per cycle versus the energy stored in the cavity. Very large $Q$ factors in the vicinity of SIPs originate from the combination of the frozen mode regime and the common slow-wave resonance \cite{figotin_slow_2011}. Nevertheless, at EPDs other than the SIP (i.e., the DBE), systems can be highly mismatched to the termination impedance of most loads. This phenomenon stems from the Floquet-Bloch impedance \cite{othman_theory_2016} in a multi TL, and causes an EPD-exhibiting structure to act as an isolated cavity. This is especially true for the case of DBEs \cite{figotin_gigantic_2005}. The ASOW does not behave as a very high-$Q$ resonator at the SIP frequency as the frozen mode regime is not a cavity resonance \cite{ figotin_slow_2011,li_frozen_2017}.

The $Q$ factor, however, does depend on the particular design of the SIP and its Bloch impedance. In the following we calculate the quality factor as  

\begin{equation}
\ensuremath{Q=\frac{\omega_{res}\tau_{g}}{2}}\label{eq:Quality Factor}
\end{equation}

that provides a very good approximation for high quality factors \cite{nada_theory_2017}. Here $\omega_{res}$ is the SIP resonance (the angular frequency corresponding to the closest peak to the SIP frequency) and $\tau_g$ is the group delay at that frequency. As the range of frequencies we operate in is small, the resonant frequency $\omega_{res}$ is approximately the same for all the group delay peaks in Fig. \ref{fig:Group-Delay-=00005Bs=00005D}. The group delay is calculated as the negative of the derivative of the phase of the transfer function with respect to the angular frequency, i.e.,

\begin{equation}
\ensuremath{\tau_{g}=-\frac{\partial\angle T_{f}}{\partial\omega}}. \label{eq: Group Delay}
\end{equation}

Figure \ref{fig:Group-Delay-=00005Bs=00005D} shows the group delay versus angular frequency for structures with different number of unit cells, $N$.

It is normalized by the baseline delay 
\\
\begin{equation}
    \tau_{0} = N\frac{n_w}{c}\left( 2\pi R + 2(\alpha'+\alpha)R\right)=N \tau'_{0}
\label{eq:BaselineGroupDelay}
\end{equation}
\\
that occurs in a finite-length structure with the same length as the ASOW, without considering the couplings (i.e., without frozen mode). For the SIP-exhibiting ASOW from Section \ref{ch:Stationary Inflection Point}, which had $R = 10$ $\upmu \text{m}$, $\alpha = 66.02 ^{\circ} $, $\alpha' = 56.18 ^{\circ} $ , we have $\tau'_{0}=0.83$ ps. In Figure \ref{fig:Group-Delay-=00005Bs=00005D} we can see the normalized group delay. As expected, for frequencies below the SIP frequency,  $\tau_g$ approximates $\tau_0$, although it does not quite reach that low value because $\tau_0$ does not take into account the resonant paths enabled by the existence of the coupling points. For frequencies above the RBE frequency, which is the frequency at which the ASOW exhibits an RBE, $\tau_g \to 0$, as there is no propagation through the waveguide and the field experiences an exponential decay while propagating inside the waveguide because of the bandgap in the dispersion diagram in Fig. 3.

\begin{figure}[H]
\centering
    \includegraphics[width = 0.48\textwidth]{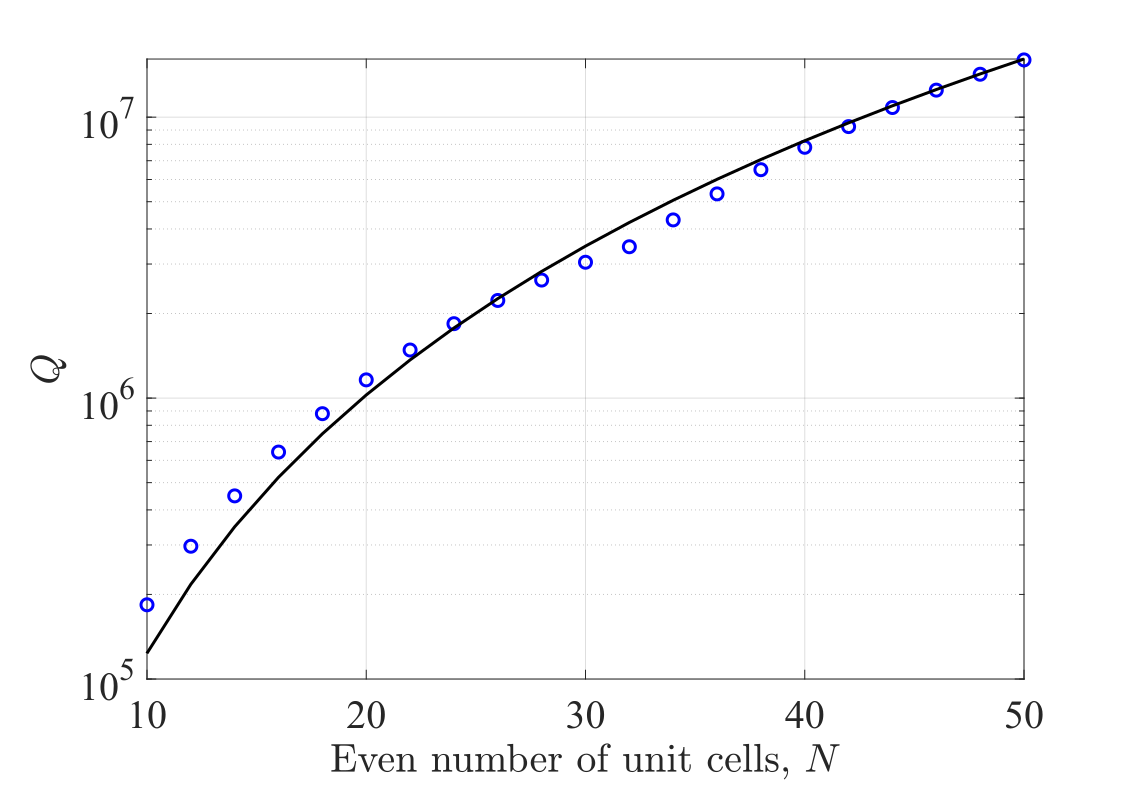}
\hfill
    \centering
    \includegraphics[width = 0.48\textwidth]{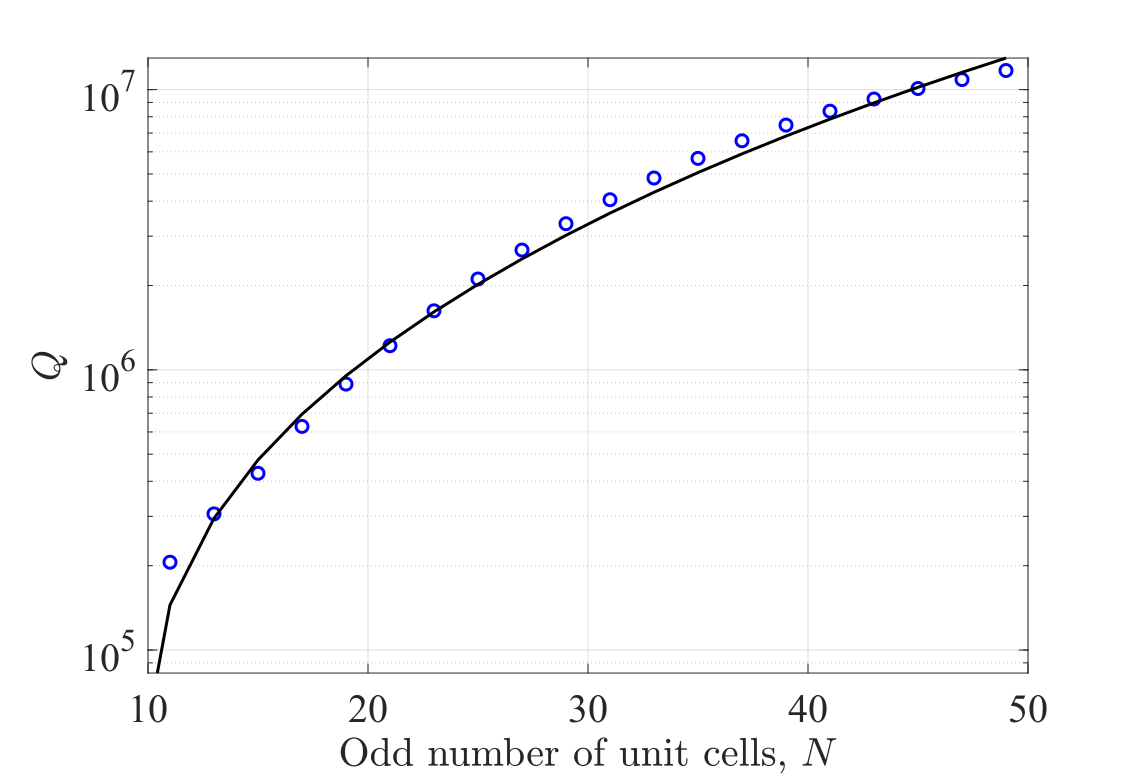}
\caption{Quality factor versus number of unit cells $N$ of the finite-length structure for (a) even-number $N$, where $Q$ evolves as $b_{e}N^3$, and (b) odd-number $N$, where $Q$ evolves as $b_{o}N^3$, with $b_e \neq b_o$.}
\label{fig:QvsN}
\end{figure}

In Figure \ref{fig:QvsN} we plot the quality factor $Q$ versus the number of unit cells $N$ of the ASOW in Fig. \ref{fig:SOW-TransferFunction}. The two plots are for even (a) and odd (b) numbers $N$.  In both cases,  $Q$ grows with the number of unit cells following the trend $Q \propto b_{e,o} N^3$ for large $N$ \cite{figotin_slow_2011}. The proportionality constants $b_e$ and $b_o$ for the even and odd $N$ cases are different from each other, with $b_e = 128.9$ and $b_o= 99.8$. The fitting curve shown in figure \ref{fig:QvsN} is  $Q = 128.9 \ N^3 - 5354$ for an even number of unit cells. For ASOWs with an odd number of unit cells, $Q = 99.8 \ N^3 + 3.2 \times 10^4 \ N  - 3.4 \times 10^5$. Both fittings are done with $N \in [20, 50]$. The high quality factors in the figures in Fig. \ref{fig:QvsN} occur because the model of the ASOW does not take into account radiation or scattering losses.

Despite the growing trend of $Q$ with $N$, the frequency at which the $Q$ is maximum does not necessarily get monotonically closer to the SIP frequency, as shown in Figure \ref{fig:Group-Delay-=00005Bs=00005D} looking at the group delay peaks. Moreover, for a relatively small number of unit cells, with $N \in [10,20]$, ASOWs with an even number of unit cells have a higher $Q$ than ASOWs with odd $N$ of comparable length, suggesting a stronger cavity-like behavior for ASOWs with even $N$. This is seen in Fig. \ref{fig:QvsN}. For larger $N$, this difference disappear.

Figure \ref{fig:Group-Delay-=00005Bs=00005D} shows the normalized group delay peaks in the vicinity of the SIP frequency and near the RBE frequency, indicating that $Q$ is higher near  EPD frequencies.
As mentioned before, the $Q$ around the SIP frequency grows as: $Q_{SIP} = b_{SIP}N^3$. Note that also for the resonances near the RBE, we have the asymptotic trend $Q_{RBE} = b_{RBE}N^3$ as discussed in  \cite{figotin_slow_2011, figotin_gigantic_2005}. 
For the ASOW considered here, the $Q$ in the vicinity of the SIP frequency is comparable to the $Q$ in the vicinity of the RBE. This occurs even though the SIP displays a frozen mode regime and has a higher degeneracy order than the RBE. 

As the SIP exhibits high transmittance, it allows a balance between the dramatic enhancement of the field amplitudes associated with an exceptional point and the low coupling to external waveguides due to mismatch. A high level of mismatch is instead typically found in DBEs, which have a higher quality factor scaling law  \cite{nada_theory_2017}. In \cite{gutman_stationary_2011}, it is shown that SIP-exhibiting structures have a high coupling coefficient, with a significant part of the incident light being transmitted into the frozen mode regime. This feature reduces the $Q$ factor of the structure and the cavity-like properties that, instead, band edges usually exhibit. As such, SIP-exhibiting structures can be devised to realize unidirectional lasers \cite{ramezani_unidirectional_2014} that are otherwise not suitable with waveguides with an even-order EPD, such as an RBE or a DBE \cite{veysi_laser_2018}, which are used to form high $Q$ cavities with low transmittance.

\section{Conclusion}
\label{ch:Discussion}

We have demonstrated that a lossless asymmetric serpentine optical waveguide (ASOW) can support a pair of reciprocal SIPs associated with the frozen mode regime. The SIP has been obtained using the extra degree of freedom by applying a shear distortion that breaks the glide symmetry of the original symmetric SOW. Our formulation explicitly reveals that the SIP is an exceptional point of third order in a lossless/gainless waveguide. To show that, we resort to the concept of "coalescence parameter" whose vanishing value reveals the coalescence of three eigenvectors, explicitly demonstrating that the SIP is indeed a third order exceptional point of degeneracy. The study of finite-length waveguides shows the field enhancement and a large group delay at Fabry-Perot resonances near the SIP frequency. We have also studied the evolution of the transfer and reflection functions in the vicinity of the SIP, varying the length of the waveguide cavity, revealing the cubic-length scaling of the quality factor. High transmission is observed, shown by a transfer function nearing 0 dB close to the SIP frequency, with a reasonably high quality factor, allowing for matching the SIP-exhibiting structure to external devices. Periodic waveguides supporting the SIP-related frozen mode regime can be used for cavity-less light amplification and lasing, optical sensors, microwave and optical modulators and switches.

\section{Appendix A}
\label{ch:AppendixA}

\begin{figure}[H]
\centering
\includegraphics[width = 0.4\textwidth]{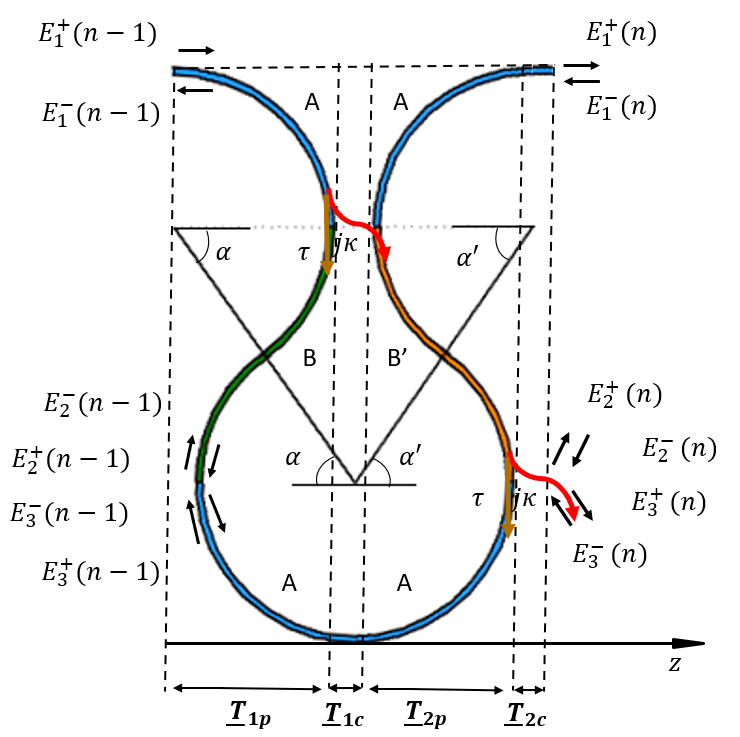}
\caption{Unit cell of the ASOW divided in subcells, which are the waveguide segments within the parallel dashed oblique lines. The subscells are modeled by the transfer matrices: $\underline{\mathbf{T}}_{1p}$, $\underline{\mathbf{T}}_{1c}$, $\underline{\mathbf{T}}_{2p}$ and $\underline{\mathbf{T}}_{2c}$. The transfer matrices $\underline{\mathbf{T}}_{ip}$, with $i = 1,2$ describe segments of the waveguide where the waves travel in three uncoupled waveguides, whereas $\underline{\mathbf{T}}_{ic}$, with $i = 1,2$, describe the coupling points, assumed to have zero thickness.}
\label{fig:UnitCellBrokenDown}
\end{figure}

In this appendix we show how to obtain the transfer matrix of the unit cell of the ASOW $\underline{\mathbf{T}}_{u}$ and the auxiliary matrix $\underline{\mathbf{T}}_{aux}$, which is akin to $\underline{\mathbf{T}}_{u}$ without modeling the second coupling point. The state vector is given in Equation (\ref{eq:StateVector}).

As the unit cell of the structure has different resonant paths, the transfer matrix for the unit cell cannot be calculated in one step. Instead, we break the unit cell into several segments shown in Figure \ref{fig:UnitCellBrokenDown}.

We call $\underline{\mathbf{T}}_{1c}$ and $\underline{\mathbf{T}}_{2c}$ the  transfer matrices that model the relations between field amplitudes on either side of the infinitesimal segments (in $z$) that include the coupling points. The transfer matrices $\underline{\mathbf{T}}_{1p}$ and $\underline{\mathbf{T}}_{2p}$ account for the phase accumulation in the segments of the unit cell.

The matrices $\underline{\mathbf{T}}_{1p}$ \& $\underline{\mathbf{T}}_{2p}$ are trivial:

\begin{equation}
    \underline{\mathbf{T}}_{1p}=\left(\begin{array}{cccccc}
e^{j\phi_{a}} & 0 & 0 & 0 & 0 & 0\\
0 & e^{-j\phi_{a}} & 0 & 0 & 0 & 0\\
0 & 0 & e^{j\phi_{b}} & 0 & 0 & 0\\
0 & 0 & 0 & e^{-j\phi_{b}} & 0 & 0\\
0 & 0 & 0 & 0 & e^{j\phi_{a}} & 0\\
0 & 0 & 0 & 0 & 0 & e^{-j\phi_{a}}
\end{array}\right)
\label{eq:T1p}
\end{equation}

and 

\begin{equation}
    \underline{\mathbf{T}}_{2p}=\left(\begin{array}{cccccc}
e^{j\phi_{a}} & 0 & 0 & 0 & 0 & 0\\
0 & e^{-j\phi_{a}} & 0 & 0 & 0 & 0\\
0 & 0 & e^{j\phi_{b}'} & 0 & 0 & 0\\
0 & 0 & 0 & e^{-j\phi_{b}'} & 0 & 0\\
0 & 0 & 0 & 0 & e^{j\phi_{a}} & 0\\
0 & 0 & 0 & 0 & 0 & e^{-j\phi_{a}}
\end{array}\right)
\label{eq:T2p}
\end{equation}

but the matrices $\underline{\mathbf{T}}_{1c}$ \& $\underline{\mathbf{T}}_{2c}$ demand a more careful consideration. They are:

\begin{equation}
    \underline{\mathbf{T}}_{1c}=\left(\begin{array}{cccccc}
0 & -j\frac{\tau}{\kappa} & \frac{j}{\kappa} & 0 & 0 & 0\\
j\frac{\tau}{\kappa} & 0 & 0 & -\frac{j}{\kappa} & 0 & 0\\
\frac{j}{\kappa} & 0 & 0 & -j\frac{\tau}{\kappa} & 0 & 0\\
0 & -\frac{j}{\kappa} & j\frac{\tau}{\kappa} & 0 & 0 & 0\\
0 & 0 & 0 & 0 & 1 & 0\\
0 & 0 & 0 & 0 & 0 & 1
\end{array}\right)
\label{eq:T1c}
\end{equation}

and

\begin{equation}
    \underline{\mathbf{T}}_{2c}=\left(\begin{array}{cccccc}
1 & 0 & 0 & 0 & 0 & 0\\
0 & 1 & 0 & 0 & 0 & 0\\
0 & 0 & 0 & -j\frac{\tau}{\kappa} & \frac{j}{\kappa} & 0\\
0 & 0 & j\frac{\tau}{\kappa} & 0 & 0 & -\frac{j}{\kappa}\\
0 & 0 & \frac{j}{\kappa} & 0 & 0 & -j\frac{\tau}{\kappa}\\
0 & 0 & 0 & -\frac{j}{\kappa} & j\frac{\tau}{\kappa} & 0
\end{array}\right).
\label{T2c}
\end{equation}

In the following we show how to obtain the transfer matrix $\underline{\mathbf{T}}_{1c}$, with $\underline{\mathbf{T}}_{2c}$ being analogously derived.
The transfer matrix $\underline{\mathbf{T}}_{1c}$ represents the infinitesimally thin (in $z$) segment with the top coupling point. As seen in Figure \ref{fig:UnitCellBrokenDown}, there is no phase accumulation at the bottom ports (identified by the field amplitudes $E_3^{\pm}$). This explains the 2x2 identity matrix at the bottom right of $\underline{\mathbf{T}}_{1c}$.

To model the change in the field amplitudes before and after the coupling point we use a 4x4 scattering matrix, which gives the outputs in terms of the inputs. For the coupling point modeled in $\underline{\mathbf{T}}_{1c}$, $z_c$, we have the following scattering matrix,

\begin{equation}
    \left(\begin{array}{c}
E_{1}^{-}(z_{c}^{-})\\
E_{2}^{-}(z_{c}^{-})\\
E_{1}^{+}(z_{c}^{+})\\
E_{2}^{+}(z_{c}^{+})
\end{array}\right)=\left(\begin{array}{cccc}
0 & \tau & 0 & -j\kappa\\
\tau & 0 & -j\kappa & 0\\
0 & -j\kappa & 0 & \tau\\
-j\kappa & 0 & \tau & 0
\end{array}\right)\left(\begin{array}{c}
E_{1}^{+}(z_{c}^{-})\\
E_{2}^{+}(z_{c}^{-})\\
E_{1}^{-}(z_{c}^{+})\\
E_{2}^{-}(z_{c}^{+})
\end{array}\right).
\label{eq:ScatteringMatrix}
\end{equation}

This 4x4 scattering matrix is transformed into the 4x4 transfer matrix embedded at the top left of the 6x6 $\underline{\mathbf{T}}_{1c}$. The transformations are \cite{nada_theory_2017}

\begin{equation}
\begin{gathered}
     \underline{\mathbf{T}}_{11}=\underline{\mathbf{S}}_{21} \ - \ \underline{\mathbf{S}}_{22} \ \underline{\mathbf{S}}_{12}^{-1} \ \underline{\mathbf{S}}_{11},\\
    \underline{\mathbf{T}}_{21}= -\underline{\mathbf{S}}_{12}^{-1} \ \underline{\mathbf{S}}_{11} ,\\
    \underline{\mathbf{T}}_{12}=\underline{\mathbf{S}}_{22} \ \underline{\mathbf{S}}_{12}^{-1}, \\
    \underline{\mathbf{T}}_{22}=\underline{\mathbf{S}}_{12}^{-1}
    \label{StoTtransformation}
\end{gathered}
\end{equation}

where each component, $\underline{\mathbf{S}}_{ij}$ and  $\underline{\mathbf{T}}_{ij}$, with $i,j = 1,2$, is a 2x2 matrix that forms the 4x4 scattering and transfer matrices, respectively. The transfer matrix, which relates the field amplitudes at the left of the coupling point ($z_c^-$) with the field amplitudes at the right of the coupling point  ($z_c^+$), is shown below:

\begin{equation}
    \left(\begin{array}{c}
E_{1}^{+}(z_{c}^{+})\\
E_{1}^{-}(z_{c}^{+})\\
E_{2}^{+}(z_{c}^{+})\\
E_{2}^{-}(z_{c}^{+})
\end{array}\right)=\left(\begin{array}{cccc}
0 & -j\frac{\tau}{\kappa} & \frac{j}{\kappa} & 0 \\
j\frac{\tau}{\kappa} & 0 & 0 & -\frac{j}{\kappa} \\
\frac{j}{\kappa} & 0 & 0 & -j\frac{\tau}{\kappa} \\
0 & -\frac{j}{\kappa} & j\frac{\tau}{\kappa} & 0 \\
\end{array}\right)\left(\begin{array}{c}
E_{1}^{+}(z_{c}^{-})\\
E_{1}^{-}(z_{c}^{-})\\
E_{2}^{+}(z_{c}^{-})\\
E_{2}^{-}(z_{c}^{-}).
\end{array}\right)
\label{eq:DirectionalCoupler}
\end{equation}

Embedding this 4x4 transfer matrix on the top left of the 6x6 transfer matrix $\underline{\mathbf{T}}_{1c}$ we obtain a full model of the infinitesimal segment with the top coupling point. 

For the transfer matrix $\underline{\mathbf{T}}_{2c}$, the coupling  occurs for the field amplitudes $E_2^\pm$ and $E_3^\pm$, so the 4x4 transfer matrix modeling the coupling point is embedded in the bottom right part of the 6x6 $\underline{\mathbf{T}}_{2c}$. As there is no change in $E_1^\pm$ (due to the aforementioned infinitesimal thickness of the modeled segment), a 2x2 identity matrix goes at the top left. The rest of the matrix is filled with zeros.

The last step to obtain the transfer matrix is to right-multiply the transfer matrix of each segment:

\begin{equation}
    \underline{\mathbf{T}}_{u}=\underline{\mathbf{T}}_{2c} \ \underline{\mathbf{T}}_{2p} \ \underline{\mathbf{T}}_{1c} \ \underline{\mathbf{T}}_{1p}.
    \label{eq:UnitCellTransferMatrixFromPieces}
\end{equation}

The full expression of $\underline{\mathbf{T}}_{u}$ is shown in Eq. (\ref{eq:TransferMatrix}). The 6x6 transfer matrix $\underline{\mathbf{T}}_{aux}$, which describes a modified unit cell without the bottom coupling to be used as a last cell, containing the outport port, is similar to the transfer matrix $\underline{\mathbf{T}}_{u}$ but without right-multiplying the matrix $\underline{\mathbf{T}}_{2c}$, as

\begin{equation}
    \underline{\mathbf{T}}_{aux}=\underline{\mathbf{T}}_{2p} \ \underline{\mathbf{T}}_{1c} \ \underline{\mathbf{T}}_{1p}
    \label{eq:TauxFromPieces}
\end{equation}

yielding

\begin{widetext}
\begin{equation}
\resizebox{0.85\textwidth}{!}{$
\underline{\mathbf{T}}_{aux} = \left(\begin{array}{cccccc}
0 & -j\frac{\tau}{\kappa} & j\frac{e^{j(\phi_{a}+\phi_{b})}}{\kappa} & 0 & 0 & 0\\
\frac{j\tau}{\kappa} & 0 & 0 & -j\frac{e^{-j(\phi_{a}+\phi_{b})}}{\kappa} & 0 & 0\\
j\frac{e^{j(\phi_{a}+\phi_{b}')}}{\kappa} & 0 & 0 & -j\frac{\tau e^{-j(\phi_{b}-\phi_{b}')}}{\kappa} & 0 & 0\\
0 & -j\frac{e^{-j(\phi_{a}+\phi_{b}')}}{\kappa} & j\frac{\tau e^{j(\phi_{b}-\phi_{b}')}}{\kappa} & 0 & 0 & 0\\
0 & 0 & 0 & 0 & e^{j2\phi_{a}} & 0\\
0 & 0 & 0 & 0 & 0 & e^{-j2\phi_{a}}
\end{array}\right)$}
\label{eq:TauxFullExpression}
\end{equation}
\end{widetext}

\section*{Acknowledgement}

This research has been conducted with the partial support from the Balsells Corporation Fellowship program that cosponsored AH from Spain to have the opportunity to conduct research and study in the United States at UC Irvine. This material is based upon work supported by the Air Force Office of Scientific Research award numbers LRIR 21RYCOR019 and FA8655-20-1-7052.


\end{document}